\documentclass[12pt]{article}

\usepackage{fullpage}
\usepackage{cite}
\usepackage{graphicx}
\usepackage{amsmath}
\usepackage{amssymb}
\usepackage{color}
\usepackage{graphics}
\usepackage{xcolor,colortbl}
\definecolor{grey}{rgb}{0.8, 0.8, 0.8}
\definecolor{lightgrey}{rgb}{0.94, 0.94, 0.94}

\allowdisplaybreaks
\title{}
\date{}

\def\beq{\begin{equation}}
\def\eeq{\end{equation}}
\def\beqa{\begin{eqnarray}}
\def\eeqa{\end{eqnarray}}

\renewcommand{\vec}[1]{\mbox{\boldmath$ #1 $}}

\newcommand{\be}{\begin{equation}}
\newcommand{\ee}{\end{equation}}
\newcommand{\bea}{\begin{eqnarray}}
\newcommand{\eea}{\end{eqnarray}}
\newcommand{\nn}{\nonumber}

\newcommand{\ord}{{\cal O}}

\newcommand{\eps}{\epsilon}

\begin{document}
\bibliographystyle{utphys}


\titlepage

\begin{flushright}
QMUL-PH-18-13\\
NIKHEF/2018-038\\
\end{flushright}

\vspace*{1.2cm}

\begin{center}
{\Large \bf On next-to-leading power threshold corrections in
  Drell-Yan production at N$^3$LO}

\vspace*{1cm} \textsc{N. Bahjat-Abbas$^a$, 
  J. Sinninghe Damst\'{e}$^{b,c}$,
  L. Vernazza$^{c}$ and
  C. D. White$^a$} \\

\vspace*{1.5cm} 

$^a$ Centre for Research in String Theory, School of Physics and
Astronomy, Queen Mary University of London, 327 Mile End Road, London
E1 4NS, UK

\vspace*{0.3cm}

$^b$ ITFA, University of Amsterdam, Science Park 904, Amsterdam, 
The Netherlands \\

\vspace*{0.3cm} 

$^c$ Nikhef, Science Park 105, NL-1098 XG Amsterdam, The Netherlands







\end{center}

\vspace*{1cm}

\begin{abstract}
  \noindent The cross-section for Drell-Yan production of a vector
  boson has been previously calculated at next-to-next-to-leading
  order, supplemented by enhanced logarithmic terms associated with
  the threshold region. In this paper, we calculate a large set of
  enhanced terms associated with the colour structure $C_F^3$ at
  N$^3$LO, for the double real emission contribution in the
  quark-antiquark channel, as an expansion around the threshold region
  up to and including the first subleading power. We perform our
  calculation using the method of regions, which systematically
  characterises all contributions according to whether the virtual
  gluon is (next-to) soft, collinear or hard in nature. Our results
  will prove useful for developing general formalisms for classifying
  next-to-leading power (NLP) threshold effects. They are also
  interesting in their own right, given that they constitute a
  previously unknown contribution to the Drell-Yan cross-section at
  ${\cal O}(\alpha_s^3)$.
\end{abstract}


\section{Introduction}
\label{intro}

The ongoing Large Hadron Collider programme, together with related
experimental facilities, necessitates the calculation of scattering
processes in perturbative quantum field theory to ever greater
precision. The state of the art in such calculations typically evolves
on two complementary fronts. Firstly, there is the calculation of
specific processes at fixed order in perturbation theory (including
both QCD and electroweak corrections). Secondly, there is the
inclusion of successive infinite towers of kinematically enhanced
contributions, and the matching of these so-called {\it resummed}
predictions with fixed order results. The state of the art for most
processes of interest is next-to-leading order (NLO) in perturbation
theory, supplemented by next-to-next-to-leading logarithmic (NNLL)
resummed contributions. A few processes are known beyond this order,
and in this paper we focus on inclusive quantities in the production
of heavy particles, which depend on a single ratio $\xi$ of kinematic
scales, such that $\xi\rightarrow 0$ at threshold. Examples include
the Drell-Yan production of a vector boson, which is currently known
to
NNLO~\cite{Drell:1970wh,Altarelli:1979ub,Hamberg:1990np,Hamberg2002403,Matsuura:1988sm,Matsuura:1990ba,Matsuura:1989sm,Matsuura:1988nd},
and the closely related process of Higgs boson production via
gluon-gluon fusion, which has been calculated up to an impressive
N$^3$LO~\cite{Dawson:1991zj,Anastasiou:2002yz,Harlander:2002wh,Anastasiou:2013srw,Anastasiou:2013mca,Anastasiou:2014vaa,Anastasiou:2014lda,Anastasiou:2015ema}
in the large top mass limit. The differential cross-section in QCD for
these and other single-scale quantities assumes the generic form \beq
\frac{d \sigma}{d \xi} \, = \, K_{\rm ew} \left( 4 \pi \alpha_s
\right)^{n_0} \sum_{n = 0}^{\infty} \left( \frac{\alpha_s}{\pi}
\right)^n \sum_{m = 0}^{2 n - 1} \left[ \, c_{n m}^{(-1)} \left(
  \frac{\log^m \xi}{\xi} \right)_+ + \, c_{n}^{(\delta)} \,
  \delta(\xi) + \, c_{nm}^{(0)} \, \log^m \xi + \ldots \, \right] ,
\label{thresholddef}
\eeq where $K_{\rm ew}$ collects electroweak coupling and
normalisation factors, $\alpha_s=g_s^2/(4\pi)$ is the strong coupling,
and $n_0$ denotes the power of the strong coupling in the Born
interaction. Commencing at NLO, each order in $\alpha_s$ is
accompanied by a series of divergent contributions as the threshold
variable tends to zero, associated with QCD radiation that is soft and
/ or collinear with the hard particles in the underlying scattering
process. The first set of terms in the square bracket in
eq.~(\ref{thresholddef}) constitutes the leading power (LP) in the
threshold variable $\xi$, which mixes with the second set of terms,
that originates also from purely virtual corrections. The third set of
terms is next-to-leading power (NLP) in a systematic expansion in
$\xi$, and formally divergent as $\xi\rightarrow 0$, albeit integrably
so. Finally, the ellipses in eq.~(\ref{thresholddef}) denotes higher
power corrections in $\xi$ which vanish at threshold.

The practical significance of threshold contributions is well-known,
and a variety of approaches exist for resumming LP terms to all orders
in perturbation
theory~\cite{Sterman:1987aj,Catani:1989ne,Korchemsky:1993xv,Kidonakis:1997gm,Contopanagos:1997nh,Forte:2002ni,Becher:2006nr}
in order to obtain meaningful comparisons of theory with data. In
recent years, the NLP terms in eq.~(\ref{thresholddef}) have also
received a great deal of attention, for a number of reasons. Firstly,
they can dominate the theoretical uncertainty in the threshold region
once the first few powers of LP logarithms have been resummed (see
e.g.~\cite{Kramer:1996iq}, and~\cite{Herzog:2014wja} for a more recent
discussion). Secondly, the origin and general structure of NLP terms -
including whether or not they share similar universality properties
with their LP counterparts - is an interesting problem of quantum
field theory in its own right. Thirdly, the classification of NLP
contributions in cross-sections is closely
related~\cite{White:2014qia} to the study of so-called {\it
  next-to-soft theorems}, which have been explored in both a gauge
theory~\cite{Casali:2014xpa} and gravitational
context~\cite{Gross:1968in,White:2011yy,Cachazo:2014fwa} due to their
intriguing relation with asymptotic symmetries. 

Whether or not a general resummation prescription exists for NLP terms
is still an open question, that has been explored using an assortment
of methods~\cite{Laenen:2008ux,Laenen:2008gt,
  Laenen:2010uz,Bonocore:2014wua,Bonocore:2015esa,Bonocore:2016awd,Moch:2009mu,
  Moch:2009hr,Soar:2009yh,Almasy:2010wn,Presti:2014lqa,deFlorian:2014vta,
  Grunberg:2009yi,Grunberg:2009vs,Larkoski:2014bxa,Kolodrubetz:2016uim,
  Moult:2016fqy,Boughezal:2016zws,Moult:2017rpl,Chang:2017atu,Feige:2017zci,
  Gervais:2017yxv,Gervais:2017zky,Beneke:2017ztn,Beneke:2017mmf}, some
of them building upon the earlier work of
refs.~\cite{Low:1958sn,Burnett:1967km,DelDuca:1990gz}. In order to
further develop and test such formalisms, it is crucial to have
detailed theoretical data - namely, explicit results for threshold
logarithms up to NLP power in specific processes. Furthermore, it is
extremely useful to classify separately contributions to each
individual NLP term that come from real or virtual radiation that is
soft and / or collinear (or hard, in the case of multiple
emissions). Drell-Yan production offers a particularly clean testing
ground in this regard, given that all threshold logarithms associated
with purely real radiation are manifestly (next-to-) soft in origin
(see e.g.~\cite{Forte:2002ni}). Virtual gluons, however, can indeed be
collinear with one of the incoming parton legs, as well as hard or
soft, thus leading to a nontrivial structure of threshold
logarithms. A convenient way to classify each individual contribution
is to carry out the integration over virtual momenta using the {\it
  method of regions}~\cite{Beneke:1997zp,Pak:2010pt,Jantzen:2011nz},
which explicitly separates out the modes of the loop momentum into
non-overlapping soft, collinear or hard configurations. This method
was heavily used in the calculation of the total cross-section for
Higgs boson production at
N$^3$LO~\cite{Anastasiou:2014vaa,Anastasiou:2014lda}, and was also
used in ref.~\cite{Bonocore:2014wua} to reanalyse the 1-real,
1-virtual contribution to the NNLO Drell-Yan cross-section, first
calculated in
refs.~\cite{Hamberg:1990np,Hamberg2002403,Matsuura:1988sm,Matsuura:1990ba,Matsuura:1989sm,Matsuura:1988nd},
to obtain the contribution associated with each separate virtual
region. This data proved essential when deriving a factorisation
formula for next-to-soft
effects~\cite{Bonocore:2015esa,Bonocore:2016awd}, which generalises
the well-known soft-collinear factorisation formula at LP (see
e.g. ref.~\cite{Dixon:2008gr}), and which may pave the way for a NLP
resummation formalism (see
refs.~\cite{Larkoski:2014bxa,Kolodrubetz:2016uim,
  Moult:2016fqy,Boughezal:2016zws,Moult:2017rpl,Chang:2017atu,Feige:2017zci,Beneke:2017ztn,Beneke:2017mmf}
for an alternative approach based on effective field theory).

Reference~\cite{Bonocore:2014wua} focused specifically on abelian-like
contributions to the $q\bar{q}$ initial state, which in QCD are
associated with the colour structure $C_F^n$ at ${\cal
  O}(\alpha_s^n)$. At any given order, such terms are amongst the most
complicated in terms of the number of different NLP effects that
underly their structure. Furthermore, the development of factorisation
formulae and / or resummation prescriptions for threshold corrections
can be made systematically simpler by beginning with the abelian-like
theory (as in
refs.~\cite{DelDuca:1990gz,Bonocore:2014wua,Bonocore:2015esa,Bonocore:2016awd}),
before generalising to the non-abelian case. We will thus restrict
ourselves to abelian-like contributions in this paper, but our aim is
to extend the classification of threshold contributions, up to NLP in
the threshold variable, to diagrams involving one virtual gluon and
two real emissions. As in ref.~\cite{Bonocore:2014wua}, the presence
of the virtual gluon means that there are non-trivial regions to
analyse. Furthermore, the results will have a direct bearing on how to
generalise the factorisation formula of
refs.~\cite{Bonocore:2015esa,Bonocore:2016awd} to include the effects
of more than one gluon emission, which is clearly a necessary
component for resummation. Although this is our main motivation, it
should be stressed that the results of this paper constitute part of
the Drell-Yan cross-section at N$^3$LO, which is not yet known,
although leading power threshold terms have been previously evaluated
in refs.~\cite{Li:2014afw,Li:2014bfa,Ahmed:2014cla}. 

The structure of our paper is as follows. In
section~\ref{sec:outline}, we review necessary facts regarding
Drell-Yan production, and outline the various steps used in our
calculation. In section~\ref{sec:results}, we present results for the
abelian-like contribution to the Drell-Yan $K$ factor, before
discussing their structure. We conclude in
section~\ref{sec:conclude}. Some technical details are contained in
the appendices.

\section{Outline of the calculation}
\label{sec:outline}

\subsection{Drell-Yan production}
\label{sec:DY}
In this section, we review some necessary facts about the Drell-Yan
process, and the method of regions, that will be needed for what
follows. Throughout, we focus on the quark-antiquark Drell-Yan
production of a colour singlet vector boson, corresponding to the LO
process
\begin{equation}
q(p)+\bar{q}(\bar{p})\rightarrow V(Q).
\label{DYLO}
\end{equation}
For our purposes, we may take $V$ to be an off-shell photon, and let
$e_q$ denote the electromagnetic charge of the incoming quark. We
further define the variable
\begin{equation}
z=\frac{Q^2}{s}, 
\label{zdef}
\end{equation}
where $Q^2$ is the virtuality of the vector boson, and
$s=(p+\bar{p})^2$ the squared centre of mass energy. At leading order,
$z=1$, such that the cross-section may be written
\begin{equation}
\sigma^{(0)}=\sigma_0 \delta(1-z),
\label{sigma0def}
\end{equation}
where
\begin{equation}
\sigma_0=\frac{e_q^2\pi(1-\epsilon)}{N_c s},
\label{sigma0def2}
\end{equation}
and $N_c$ is the number of colours. At higher orders, one has $0\leq
z\leq 1$, such that the upper limit corresponds to threshold
production. We may then define the $K$ {\it factor}
\begin{equation}
\left(\frac{\alpha_s}{4\pi}\right)^n\,
K^{(n)}(z)=\frac{1}{\sigma_0}\frac{d\sigma^{(n)}(z)}{dz},
\label{Kdef}
\end{equation}
where the right-hand side contains the differential cross-section at
${\cal O}(\alpha_s^n)$. The complete $K$ factor for Drell-Yan
production, including all partonic channels and full $z$ dependence,
has been previously calculated up to NNLO
($n=2$)~\cite{Altarelli:1979ub,Hamberg:1990np,Hamberg2002403,Matsuura:1988sm,Matsuura:1990ba,Matsuura:1989sm,Matsuura:1988nd},
and leading power threshold contributions at N$^3$LO have been
evaluated in refs.~\cite{Li:2014afw,Li:2014bfa,Ahmed:2014cla}. At any
given order, one must include the effects of additional radiation,
that may be real or virtual. Reference~\cite{Bonocore:2014wua}
reanalysed the 1-real, 1-virtual contribution to $K^{(2)}$ (for the
$q\bar{q}$ channel), up to the first subleading power in a threshold
expansion about $z=1$. In this limit, the $K$ factor assumes a form
similar to eq.~(\ref{thresholddef}), containing plus distributions and
logarithms of the threshold variable $\xi=1-z$. As discussed in the
introduction, ref.~\cite{Bonocore:2014wua} focused on all
contributions up to next-to-leading power (NLP) in $\xi$, that are
proportional to the colour factor $C_F^2$, where $C_F$ is the
quadratic Casimir in the fundamental representation. Such
contributions are similar to those one would obtain in an Abelian
theory, upon replacing $g_s C_F$ with the relevant electromagnetic
charge of the quark, and the aim of ref.~\cite{Bonocore:2014wua} was
to classify the precise origin of all such contributions, according to
whether the virtual gluon is hard, soft or collinear with one of the
incoming (anti-)quarks. Here, we carry out a similar analysis for the
case of one virtual gluon, and two real emissions. This contributes to
the N$^3$LO factor $K^{(3)}(z)$, and the virtual gluon has a number of
non-trivial momentum regions that give rise to NLP terms.

\begin{figure}
\begin{center}
\scalebox{0.6}{\includegraphics{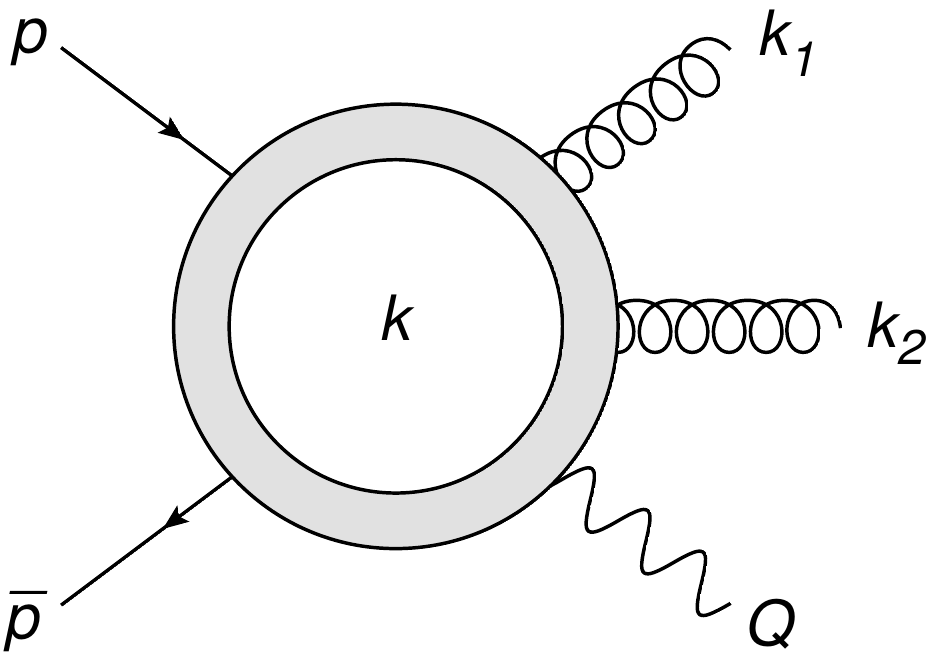}}
\caption{Contribution to the Drell-Yan process at N$^3$LO, consisting
  of two real gluon emissions dressing the one-loop amplitude. The
  latter involves a loop momentum $k$.}
\label{fig:loopdiag}
\end{center}
\end{figure}

\begin{figure}[t]
\begin{center}
\scalebox{0.5}{\includegraphics{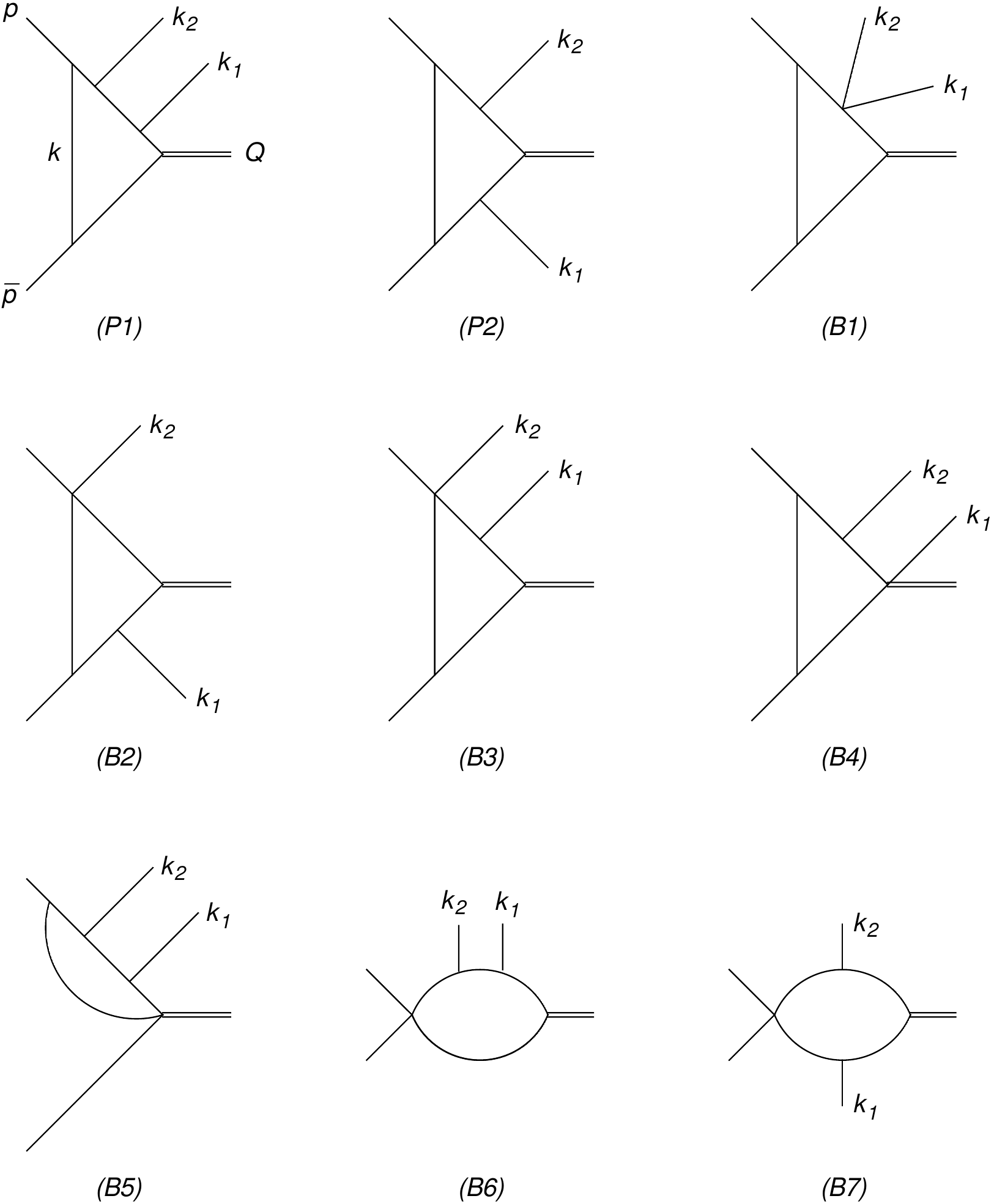}}
\caption{Pentagon  $(P_i)$ and box $(B_i)$ scalar master 
 diagrams that contribute to eq.~(\ref{sigma32r1v}).}
\label{fig:masters}
\end{center}
\end{figure}

The amplitude we consider is shown schematically in
figure~\ref{fig:loopdiag}, and corresponds to the process
\begin{equation}
q(p)+\bar{q}(\bar{p})\rightarrow V^*(Q)+g(k_1)+g(k_2)
\label{qqgg}
\end{equation}
at one-loop order.
Labelling this by ${\cal A}_{\rm 2r,1v}$, its contribution to the
differential cross-section occurs through interference with the pure
two real emission amplitude ${\cal A}_{\rm 2r}$:
\begin{align}
\frac{d\sigma_{2r, 1v}}{dz}=\frac{1}{4N_c^2}\frac{1}{2s}
2{\rm Re}\bigg[&\int \frac{d^d k}{(2\pi)^d}
\int d\Phi^{(3)}\delta \left(z-\frac{Q^2}{s}\right) \notag \\ &\times
{\cal A}_{\rm 2r, 1v}(p,\bar{p},k_1,k_2,k)\,
{\cal A}^\dag_{2r}(p,\bar{p},k_1,k_2)\bigg],
\label{sigma32r1v} 
\end{align}
where the prefactors originate from colour / spin averaging and the
Lorentz-invariant flux factor, we work in $d=4-2\epsilon$ spacetime
dimensions throughout, and $d\Phi^{(3)}$ is the differential phase
space for the 3-body final state. There are 48 distinct Feynman
diagrams that contribute to the abelian-like one-loop amplitude (where
we define abelian-like diagrams to be those that contribute to the
$C_F^3$ colour structure in the cross section, thereby also excluding
diagrams with a fermion loop). We have generated all such diagrams
using \texttt{QGRAF}\cite{Nogueira:1991ex}, and subsequently used
\texttt{Reduze}~\cite{Studerus:2009ye,vonManteuffel:2012np} (version
2) to construct the interference term appearing in
eq.~(\ref{sigma32r1v}). At this stage, one must carry out the
integration over the loop momentum $k$ appearing in
eq.~(\ref{sigma32r1v}) and figure~\ref{fig:loopdiag}. To this end, we
also use \texttt{Reduze} to reduce the one-loop integration to a set
of scalar master integrals, using integration by parts
identities. These integrals may themselves be represented as scalar
Feynman diagrams with topologies of increasing complexity. The box and
pentagon master diagrams are shown in figure~\ref{fig:masters}, where
the simpler bubbles and triangles are omitted for brevity.

As stated above, the aim of our paper is to classify the structure of
the $K$ factor up to NLP terms in the threshold expansion. We must
then consider each master integral, and elucidate its corresponding
contribution to threshold behaviour, according to whether the loop
momentum is hard, soft or (anti-)collinear to one of the incoming
partons. Here we follow the standard approach of the {\it method of
  regions}~\cite{Beneke:1997zp,Pak:2010pt,Jantzen:2011nz}, which we
describe more fully in the following section.

\subsection{The method of regions}
\label{sec:regions}

\begin{figure}
\begin{center}
\scalebox{0.56}{\includegraphics{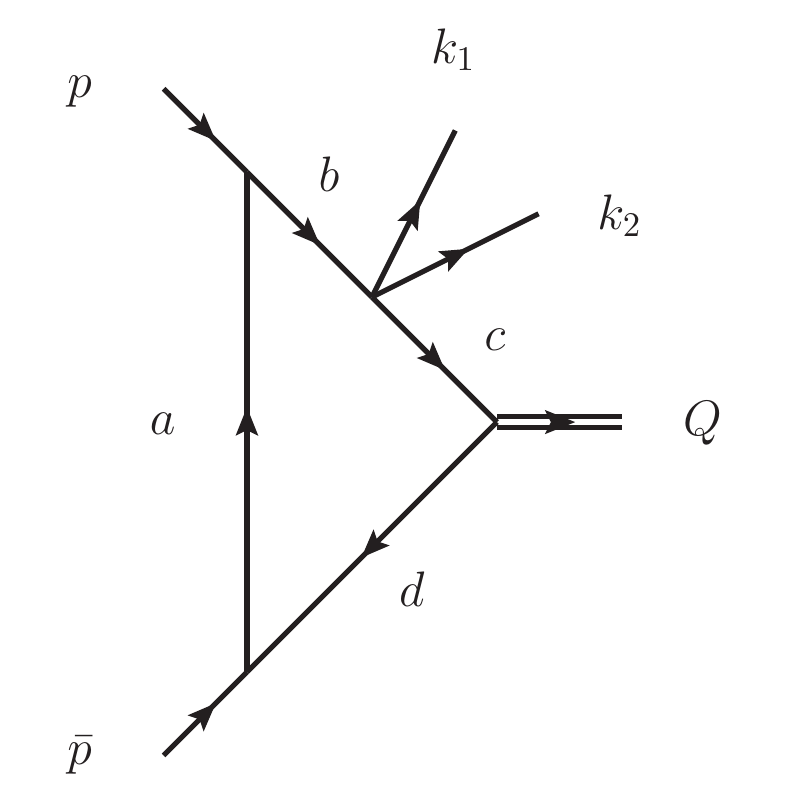}}
\caption{A particular master diagram, with internal lines labelled for
  convenience.}
\label{BoxRegions}
\end{center}
\end{figure}

In the method of regions, singular parts of integrals in perturbative
amplitudes are partitioned, according to physical criteria on the loop
momenta. In the case of the threshold expansion considered in this
paper, it is possible to separate completely the singular behaviour
into {\it non-overlapping} regions, whose individual contributions
reconstruct the full integral (itself expanded about the threshold
limit) when summed. As an example, consider the diagram $(B_1)$ of
figure~\ref{fig:masters}, where we have associated the loop momentum
$k$ with a particular internal line. One may expand this momentum in a
{\it Sudakov decomposition}
\begin{equation}
k^\mu=\frac{1}{2}(n_-\cdot k)\,n_+^\mu+\frac{1}{2}(n_+\cdot k)\,n_-^\mu
+k_\perp^\mu\equiv k_+\,n_+^\mu + k_-\,n_-^\mu+k_\perp^\mu,
\label{Sudakov}
\end{equation}
where we have defined dimensionless lightlike vectors
\begin{equation}
n_+^\mu =\frac{2}{\sqrt{s}} \,p^\mu,\quad 
n_-^\mu=\frac{2}{\sqrt{s}} \,\bar{p}^\mu,\quad n_-\cdot n_+=2
\label{n+-def}
\end{equation}
in the directions of the incoming particles, as well as the vector
$k_\perp$ transverse to the beam direction i.e. such that
\begin{equation}
k_\perp\cdot n_-=k_\perp\cdot n_+=0.
\label{kperpdots}
\end{equation}
Denoting the Sudakov components of the loop momentum via
$k^\mu=(k_+,\vec{k}_\perp,k_-)$, we may define the various regions by
different scaling behaviours of these components. That is, one may
introduce a book-keeping parameter $\lambda\sim \sqrt{1-z}$, such that the
regions we need to consider are given by momenta of the form
\beqa
  & & {\rm Hard:} \quad k\sim \sqrt{s} \left( 1, 1, 1 \right) \, ; \quad \;\;\;\;
  {\rm Soft:} \quad k\sim \sqrt{s} \left( \lambda^2, \lambda^2, \lambda^2 \right) \, ;
  \nonumber  \\
  & & {\rm Collinear:} \quad k\sim \sqrt{s} \left( 1, \lambda, \lambda^2 \right) \, ; \quad
  {\rm Anti-collinear:} \quad k\sim \sqrt{s} \left( \lambda^2, \lambda, 1 \right) \, ,
\label{scalings}
\eeqa where the terms {\it collinear} and {\it anti-collinear} denote
collinearity with respect to $p$ and $\bar{p}$ respectively. In any
given (scalar) master integral, the denominators can be systematically
expanded in $\lambda$ in each region, keeping the first subleading
power where necessary to achieve NLP order in the final expression for
the $K$ factor. The integral in each region can then be carried out, and
the results from all regions added together to reproduce, in
principle, the threshold expansion of the full integral. Note that
these are not the only possible scalings: in principle, it is also
possible to consider momenta scaling as
\begin{align*}
&{\rm Semi\mbox{-}hard}:\qquad k\sim \sqrt{s} (\lambda,\lambda,\lambda); \quad
{\rm Hard\mbox{-}collinear}:\quad  k\sim \sqrt{s} (1,\sqrt{\lambda},\lambda);
\notag\\
&{\rm Ultra\mbox{-}collinear}:\quad
 k\sim \sqrt{s} (1,\lambda^2,\lambda^4) 
\end{align*}
and so on. It is possible, however, to show that the only regions
relevant for the threshold expansion are the hard, (anti-)collinear
and soft regions defined by the scalings of eqs. (\ref{scalings}). All
other regions give scaleless integrals, which vanish in dimensional
regularisation, such that we may discard them in the following. By
definition, the incoming momenta are (anti-)collinear:
\be
p \sim \sqrt{s}(1,0,0), \qquad \qquad \bar p \sim \sqrt{s}(0,0,1),
\ee
while the gluon momenta are soft, i.e.
\be
k_1 \sim  k_2 \sim \sqrt{s}(\lambda^2,\lambda^2,\lambda^2).
\ee
There is an interesting subtlety in the above procedure, if one wants
to be sure of having characterised {\it all} possible regions of a
given master integral. Before the region expansion, a given master
integral possesses a symmetry under shifts of the loop momentum, such
that one may associate the loop momentum $k$ with an arbitrary
internal line of the master diagram. However, the decomposition of $k$
into regions breaks Lorentz invariance, leading to a violation of the
shift symmetry. It may then be the case that particular choices of $k$
are such that one cannot unambiguously identify all possible
regions. To illustrate this point, let us consider diagram $(B1)$ of
figure~\ref{fig:masters}, which we redraw in
figure~\ref{BoxRegions} so as to label the internal lines in what
follows.
In this particular case, a na\"{i}ve choice of loop momentum will
indeed lead to an important region being missed. Furthermore, this is
a problem that arises for the first time at N$^3$LO, due to requiring
the presence of a virtual gluon, and two real emissions. Given that
this problem does not seem to have been spelled out in related
calculations in the literature (e.g.~\cite{Anastasiou:2014lda}), we
believe it is instructive to examine this example in detail here.

We consider the expansion in regions of the box integral represented
in figure \ref{BoxRegions}. The integral is defined as
\be\label{box-def} I = \int [dk] \frac{1}{D_a \, D_b \, D_c \, D_d},
\ee 
where $D_i$ represents the propagator associated with line $i$ in figure
\ref{BoxRegions}, and we have introduced the convenient notation
\be
\int [dk] \equiv \frac{e^{\eps \gamma_{\rm E}}}{(4\pi)^{\eps}} \, \mu^{2\eps}_{\rm \overline{MS}} 
\int \frac{d^dk}{(2\pi)^d},
\ee
where $d = 4-2\eps$, and
$\mu_{\rm \overline{MS}} = \mu \, e^{-\gamma_{\rm E}/2} (4\pi)^{1/2}$.
Choosing the loop momentum $k$ to correspond to line $a$ seems
natural, because in this way the regions are directly associated with
having a hard, collinear or soft ``gluon'' exchange in the loop, which
should be easily interpreted in the context of an effective field
theory containing soft and collinear gluons. We can then define the
denominators
\bea \label{box-param1} \nn
D_a &=& k^2, \\ \nn
D_b &=& (k+p)^2 = k^2 +2k\cdot p, \\ \nn
D_c &=& (k+p-k_1 - k_2)^2 = k^2 +2k\cdot p - 2k\cdot (k_1+k_2)  - 2p\cdot (k_1+k_2) + 2k_1 \cdot k_2, \\ 
D_d &=& (k-\bar p)^2 =  k^2 -2k\cdot \bar p, 
\eea
and expand the loop momentum $k$ in regions using the Sudakov
decomposition of eq.~(\ref{Sudakov}). One obtains (writing $a\cdot b
\equiv a b$ in places so as to compactify expressions),
\bea \label{box-param1-LC} \nn
D_a &=& k^2, \\ \nn
D_b &=& k^2 +\sqrt{s} \, n_+ k, \\ \nn
D_c &=& k^2 +\sqrt{s}\, n_+ k 
- n_- k \, n_+(k_1 + k_2) 
- n_+ k \, n_-(k_1 + k_2) 
- k_{\perp}  (k_1+ k_2)_{\perp} \\ \nn 
&& - \sqrt{s} \, n_+(k_1+k_2) + 2k_1  k_2, \\ 
D_d &=&  k^2 - \sqrt{s} \, n_- k. 
\eea
The scaling in $\lambda$ of the various terms 
in the different regions is provided in
tables \ref{tabscaling1} and  \ref{tabscaling2}.
\begin{table}
\begin{center}
\begin{tabular}{|c|c|c|}
\hline 
$D_b$ & $k^2$  & $ \sqrt{s} \, n_+ k$  \\ \hline
(h)       & \cellcolor{grey}1  & \cellcolor{grey}1  \\ \hline
(c)       &  \cellcolor{grey}$\lambda^2$ & \cellcolor{grey}$\lambda^2$ \\ \hline
($\rm \overline{c}$)  &  $\lambda^2$ & \cellcolor{grey}1  \\ \hline
(s) & $\lambda^4$ & \cellcolor{grey}$\lambda^2$  \\ \hline
\end{tabular} \qquad \qquad 
\begin{tabular}{|c|c|c|}
\hline 
$D_d$ & $k^2$  & $- \sqrt{s} \, n_- k$  \\ \hline
(h)       & \cellcolor{grey}1  & \cellcolor{grey}1  \\ \hline
(c)       &  $\lambda^2$ & \cellcolor{grey}1 \\ \hline
($\rm \overline{c}$)  &  \cellcolor{grey}$\lambda^2$ & \cellcolor{grey}$\lambda^2$  \\ \hline
(s) & $\lambda^4$ & \cellcolor{grey}$\lambda^2$  \\ \hline
\end{tabular}
\end{center}
\caption{Scaling associated with the 
terms in the propagators $D_b$ and $D_d$,
as defined in eq.~(\ref{box-param1-LC}), where we set $s \sim 1$.
Leading terms in each region are highlighted in grey.}
\label{tabscaling1}
\end{table}
\begin{table}
\begin{center}
\begin{tabular}{|c|c|c|c|c|c|c|c|}
\hline 
$D_c$
& {\scriptsize $k^2$}  
& {\scriptsize $\sqrt{s}\, n_+ k $}
& {\scriptsize $- n_-k \, n_+(k_1 + k_2)$}
& {\scriptsize $- n_+k \, n_-(k_1 + k_2)$}
& {\scriptsize $- k_{\perp} \cdot (k_1+ k_2)_{\perp}$} 
& {\scriptsize $- \sqrt{s} \, n_+(k_1+k_2)$} 
& {\scriptsize $2k_1 \cdot k_2$}  \\ \hline
(h)       & \cellcolor{grey}1  & \cellcolor{grey}1 & $\lambda^2$   
& $\lambda^2$ & $\lambda^2$ & $\lambda^2$  & $\lambda^4$ \\ \hline
(c)       &  \cellcolor{grey}$\lambda^2$ & \cellcolor{grey}$\lambda^2$ 
& \cellcolor{grey}$\lambda^2$ & $\lambda^4$ & $\lambda^3$ & \cellcolor{grey}$\lambda^2$  
&  $\lambda^4$ \\ \hline
($\rm \overline{c}$)  &  $\lambda^2$ & \cellcolor{grey}1 
& $\lambda^4$ & $\lambda^2$ & $\lambda^3$ & $\lambda^2$ & $\lambda^4$ \\ \hline
(s) & $\lambda^4$ & \cellcolor{grey}$\lambda^2$ 
& $\lambda^4$ & $\lambda^4$ & $\lambda^4$ & \cellcolor{grey}$\lambda^2$ & 
$\lambda^4$ \\ \hline
\end{tabular}
\end{center}
\caption{Scaling associated with the 
terms in the propagators $D_c$,
as defined in eq.~(\ref{box-param1-LC}). 
Leading terms in each region 
are highlighted in grey.}
\label{tabscaling2}
\end{table}
In the following we keep 
only the leading terms for each propagator, 
thus getting the leading power contribution 
to the box integral. The hard region turns out to give
\bea\label{Bhard} \nn
I_{\rm h} &=& \int [dk] \frac{1}{k^2 \big(k^2 +\sqrt{s} \, n_+ k\big)^2
\big(k^2 - \sqrt{s} \, n_- k\big)} \\
&=&  \frac{i}{(4\pi)^2} \left(\frac{\mu_{\rm \overline{MS}}^2}{-s} \right)^{\eps} 
\frac{1}{s^2} \left( \frac{2}{\eps} - \eps \zeta_2 
- \frac{14\zeta_3}{3} \eps^2 +\ord(\eps^3)  \right),
\eea
Following the same criterion, a naive expansion in the collinear
region, assuming the scaling assigned in table \ref{tabscaling1} and
\ref{tabscaling2} gives, to leading power,
\begin{align} \label{Bcoll} \nn
I_{\rm c}&= \nn \int [dk] \frac{1}{k^2 \big(k^2 +\sqrt{s} \, n_+ k\big)
\big(k^2 +\sqrt{s} \, n_+ k - n_-k \, n_+(k_1 + k_2) - \sqrt{s} \, n_+(k_1+k_2) \big)
\big(-\sqrt{s} \, n_- k\big)}  \\ 
&= - \frac{i}{4\pi^2} \left(\frac{\mu_{\rm \overline{MS}}^2}{ \sqrt{s} \, n_+(k_1+k_2)} \right)^{\eps} 
\frac{1}{s^{3/2} \, n_+(k_1+k_2)} \left(  \frac{2}{\eps^2} - \zeta_2 - \frac{14\zeta_3}{3} \eps 
- \frac{47\zeta_4}{8} \eps^2 +\ord(\eps^3)\right).
\end{align}
Note that the hard region gives a subleading power contribution
compared to the collinear region. Within a consistent expansion to
leading power the hard region is thus zero, even if it is not
scaleless.  Furthermore, is it possible to show that integration in
the anti-collinear and soft regions give scaleless results:
\bea \label{B-cb-s} \nn
I_{\rm \bar c} &=& \int [dk] \frac{1}{k^2 \big(\sqrt{s} \, n_+ k\big)^2
\big(k^2 - \sqrt{s} \, n_- k\big)} \, = \, 0 \\ 
I_{\rm s} &=& \int [dk] \frac{1}{k^2 \big(\sqrt{s} \, n_+ k\big)
\big(\sqrt{s} \, n_+ k - \sqrt{s} \, n_+(k_1+k_2) \big)
\big(- \sqrt{s} \, n_- k\big)} \, = \, 0. 
\eea
Thus, the leading power contribution to the integral in
eq.~(\ref{box-def}) seems to be given by the collinear region in
eq.~(\ref{Bcoll}). This conclusion is erroneous, however, as an
important contribution has been missed, where the latter can be
revealed easily by shifting the loop momentum to $k' = k +p$.  As
discussed above, shift symmetry is broken by the region expansion,
such that shifting the loop momentum can lead to inequivalent regions
in general. With the new choice of loop momentum, the propagators read
\bea \label{box-param2} \nn
D_a &=& (k'- p)^2 = k'^2 - 2k'\cdot p, \\ \nn
D_b &=& k'^2, \\ \nn
D_c &=& (k'-k_1 - k_2)^2 = k'^2 - 2k'\cdot (k_1 + k_2) + 2 k_1 \cdot k_2, \\ 
D_d &=& (k'-p - \bar p)^2 =  k'^2 - 2k'\cdot (p + \bar p) + 2p\cdot \bar p,
\eea
so that applying the Sudakov decomposition of eq.~(\ref{Sudakov})
gives
\bea \label{box-param2-LC} \nn
D_a &=& k'^2 - \sqrt{s}\, n_+ k', \\ \nn
D_b &=& k'^2, \\ \nn
D_c &=& k'^2 - n_-k' \, n_+(k_1 + k_2) 
- n_+k' \, n_-(k_1 + k_2) 
- k'_{\perp} \cdot (k_1+ k_2)_{\perp} 
+ 2 k_1 \cdot k_2, \\ 
D_d &=&  k'^2 -  \sqrt{s}\, ( n_+ k'+  n_- k') + s.
\eea
The scaling of the various component in the different regions is
provided in tables \ref{tabscaling1b} and \ref{tabscaling2b}. Notice
that we label the new regions with a prime, to distinguish them from
the regions considered with the previous parameterization.
\begin{table}
\begin{center}
\begin{tabular}{|c|c|c|}
\hline 
$D_a$ & $k'^2$  & $- \sqrt{s}\, n_+ k'$  \\ \hline
(h$'$)       & \cellcolor{grey}1  & \cellcolor{grey}1  \\ \hline
(c$'$)       &  \cellcolor{grey}$\lambda^2$ & \cellcolor{grey}$\lambda^2$ \\ \hline
($\rm \overline{c}'$)  &  $\lambda^2$ & \cellcolor{grey}1  \\ \hline
(s$'$) & $\lambda^4$ & \cellcolor{grey}$\lambda^2$  \\ \hline
\end{tabular} \qquad \qquad 
\begin{tabular}{|c|c|c|c|c|}
\hline 
$D_d$ & $k'^2$  & $-  \sqrt{s}\, n_+ k'$ & $-  \sqrt{s}\, n_- k'$ & $s$  \\ \hline
(h$'$)       & \cellcolor{grey}1  & \cellcolor{grey}1 & \cellcolor{grey}1 & \cellcolor{grey}1 \\ \hline
(c$'$)       &  $\lambda^2$ &  $\lambda^2$ & \cellcolor{grey}1 & \cellcolor{grey}1 \\ \hline
($\rm \overline{c}'$)  & $\lambda^2$ & \cellcolor{grey} 1 & $\lambda^2$ & \cellcolor{grey} 1 \\ \hline
(s$'$) & $\lambda^4$ & $\lambda^2$ & $\lambda^2$ & \cellcolor{grey} 1  \\ \hline
\end{tabular}
\end{center}
\caption{Scaling associated with the 
terms in the propagators $D_a$ and 
$D_d$, as defined in eq.~(\ref{box-param2-LC}).}
\label{tabscaling1b}
\end{table}
\begin{table}
\begin{center}
\begin{tabular}{|c|c|c|c|c|c|}
\hline 
$D_c$ 
& $k'^2$ 
& $- n_-k' \, n_+(k_1 + k_2)$ 
& $- n_+k' \, n_-(k_1 + k_2) $  
&$- k'_{\perp} \cdot (k_1+ k_2)_{\perp}$  
& $2 k_1 \cdot k_2$ \\ \hline
(h$'$)       & \cellcolor{grey}1 & $\lambda^2$ & $\lambda^2$ & $\lambda^2$  & $\lambda^4$ \\ \hline
(c$'$)       &  \cellcolor{grey}$\lambda^2$ & \cellcolor{grey}$\lambda^2$ & $\lambda^4$ & 
$\lambda^3$  & $\lambda^4$ \\ \hline
($\rm \overline{c}'$)  & \cellcolor{grey} $\lambda^2$ & $\lambda^4$ & \cellcolor{grey} $\lambda^2$ 
& $\lambda^3$  & $\lambda^4$ \\ \hline
(s$'$) & \cellcolor{grey} $\lambda^4$ & \cellcolor{grey} $\lambda^4$ & \cellcolor{grey} $\lambda^4$ 
& \cellcolor{grey} $\lambda^4$  & \cellcolor{grey} $\lambda^4$\\ \hline
\end{tabular}
\end{center}
\caption{Scaling associated with the 
terms in the propagators $D_c$, 
as defined in eq.~(\ref{box-param2-LC}).}
\label{tabscaling2b}
\end{table}
It is easy to check that the new hard, collinear and anti-collinear
regions still give the same result as the old corresponding regions:
\bea\label{hard_coll_param2} \nn
I_{\rm h'} &=& \int [dk'] \frac{1}{\big(k^2 -\sqrt{s}\, n_+ k'\big) \big(k'^{\,2}\big)^2 
\big(k^2 -  \sqrt{s}\, ( n_+ k'+  n_- k') + s\big)}
\,=\, I_{\rm h}, \\ \nn
I_{\rm c'} &=& \int [dk] \frac{1}{\big(k^2 - \sqrt{s}\, n_+ k' \big) \big(k'^{\,2}\big)
\big(k'^2 - n_-k' \, n_+(k_1 + k_2)  \big)(-  \sqrt{s}\, n_- k' + s)}
\,=\, I_{\rm c}, \\
I_{\rm \bar c'} &=& \int [dk] \frac{1}{\big(- \sqrt{s}\, n_+ k' \big) \big(k'^{\,2}\big)
\big(k'^2 - n_+k' \, n_-(k_1 + k_2)  \big)(-  \sqrt{s}\, n_+ k' + s)}
\,=\, I_{\rm \bar c} = 0.
\eea
The new soft region, however, is not scaleless, and gives a new
contribution which was not present in the old parameterization:
\bea \label{soft-param2}\nn
I_{\rm s'} &=& \int [dk] \frac{1}{\big(- \sqrt{s}\, n_+ k'\big)\big(k'^{\,2}\big)
\big(k'^2 - k' \cdot (k_1 + k_2) + 2 k_1 \cdot k_2 \big) \, s} \\ 
&=& - \frac{i}{4\pi^2}\left(\frac{\mu^2}{- 2 k_1 \cdot k_2} \right)^{\eps} 
\frac{1}{s^{3/2} \, n_+(k_1+k_2)}
\left[ - \frac{1}{\eps^2} +\frac{\zeta_2}{2} + \frac{7\zeta_3}{3}\eps 
+\frac{47\zeta_4}{16} \eps^2 +\ord(\eps^3) \right]. 
\eea
In order to reconcile these results, note that the problem with the
original choice of loop momentum is that the external scales are not
well separated: both the ``collinear'' scale
$\sqrt{s} \, n_+(k_1+k_2) \sim \lambda^2$ and the ``soft'' scale
$2 k_1 \cdot k_2 \sim \lambda^4$ appear in the same propagator
$D_c$. This causes problems in the collinear region because, even if
the leading power terms in $D_c$ scales as $~\lambda^2$ (see table
\ref{tabscaling2b}), the loop integration is still over the full
domain. There is therefore a region of the integration domain in which
$k \sim - p$, so that one has
\bea\label{leading_canc}\nn
D_c|_{\rm leading \, collinear} &=& k^2 +\sqrt{s}\, n_+ k 
- n_-k \, n_+(k_1 + k_2) - \sqrt{s} \, n_+(k_1+k_2) \\ 
&\stackrel{k \sim - p}{\to}&  \sqrt{s} \, n_+(k_1 + k_2) - \sqrt{s} \, n_+(k_1+k_2)
\to 0,
\eea
i.e. the leading power terms $\sim\lambda^2$ cancel, causing the
subleading power term $2k_1 \cdot k_2 \sim \lambda^4$ to become leading.
Considering this term small in the expansion of the propagator thus leads to the wrong
analytic structure of the integral in this limit. The consequence is that
the propagator $D_c$ cannot be expanded in the collinear
region when parametrizing the loop momentum as in eq.~\eqref{box-param1}. Rather, one needs to consider a more general collinear region
``$C$'', in which the propagator $D_c$ is kept unexpanded:
\bea \label{BC}\nn
I_{\rm C} &=& \int [dk] 
\frac{1}{k^2 +\sqrt{s} \, n_+ k - 2 k\cdot (k_1 + k_2)- \sqrt{s} \, n_+(k_1+k_2) +2k_1 \cdot k_2}\\
&&\hspace{2.0cm} \cdot \, \frac{1}{k^2} \,\frac{1}{k^2 +\sqrt{s} \, n_+ k} \, \frac{1}{- \sqrt{s} \, n_- k}.
\eea
Evaluating $I_{\rm C}$ exactly and expanding at threshold 
\emph{after} integration, indeed one finds that it contains
both the contribution from the collinear and the soft region
associated with the alternative loop momentum choice of
eq.~(\ref{box-param2}):
\bea \label{B2Cres} \nn
I_{\rm C} &=& - \frac{i}{(4\pi)^2}
\frac{1}{s^{3/2} \, n_+(k_1+k_2)}\frac{\Gamma^2(-\eps)\Gamma(\eps)}{\Gamma(-2\eps)}  \\&\times&\Bigg[
\frac{1}{2}\left(\frac{\mu_{\rm \overline{MS}}^2}{-2k_1 \cdot k_2} \right)^{\eps}- \left(\frac{\mu_{\rm \overline{MS}}^2}{ \sqrt{s} \, n_+(k_1+k_2)} \right)^{\eps} 
 \Bigg]\, =\, I_{\rm s'} + I_{\rm c'}.
\eea
An independent check can be performed with the program
\texttt{Asy}~\cite{Pak:2010pt,Jantzen:2012mw}, which provides a
geometrical method to reveal the regions contributing to a given
integral. Using the program with the integral in eq.~(\ref{box-def})
reveals the existence of three non-scaleless regions, which correspond
to the hard, collinear and soft regions found within the second
parameterization of the loop momentum in eq.~(\ref{box-param2}). The
same program can be used to verify that we have captured all regions
in every other diagram.

Some readers may be wondering why the hard region exhibits infrared
singularities in the above results, which can be a common point of
confusion in the method of regions. The approach we have taken above
is to perform all required momentum scalings, and to set to zero any
integrals which remain scaleless in dimensionless regularisation. In
the soft region, expansion of the propagators changes the ultraviolet
scaling behaviour of the integral, and thus introduces (spurious)
ultraviolet divergences, whose effect is to cancel infrared
divergences associated with exchange of multiple gluons between the
incoming (anti)-quark legs, i.e. associated with the scale $s$. One
can instead choose to isolate these UV divergences and absorb them
into the hard function, and the effect of this procedure is to
transfer poles in $\epsilon$ from the hard to the soft region. Given
that this has no bearing on the final result for the $K$ factor (which
is a sum of all regions), we do not do this here. However, it should
be remembered throughout that $\epsilon$ poles appearing in the hard
region are indeed of soft origin.

\begin{figure}
\begin{center}
\scalebox{0.5}{\includegraphics{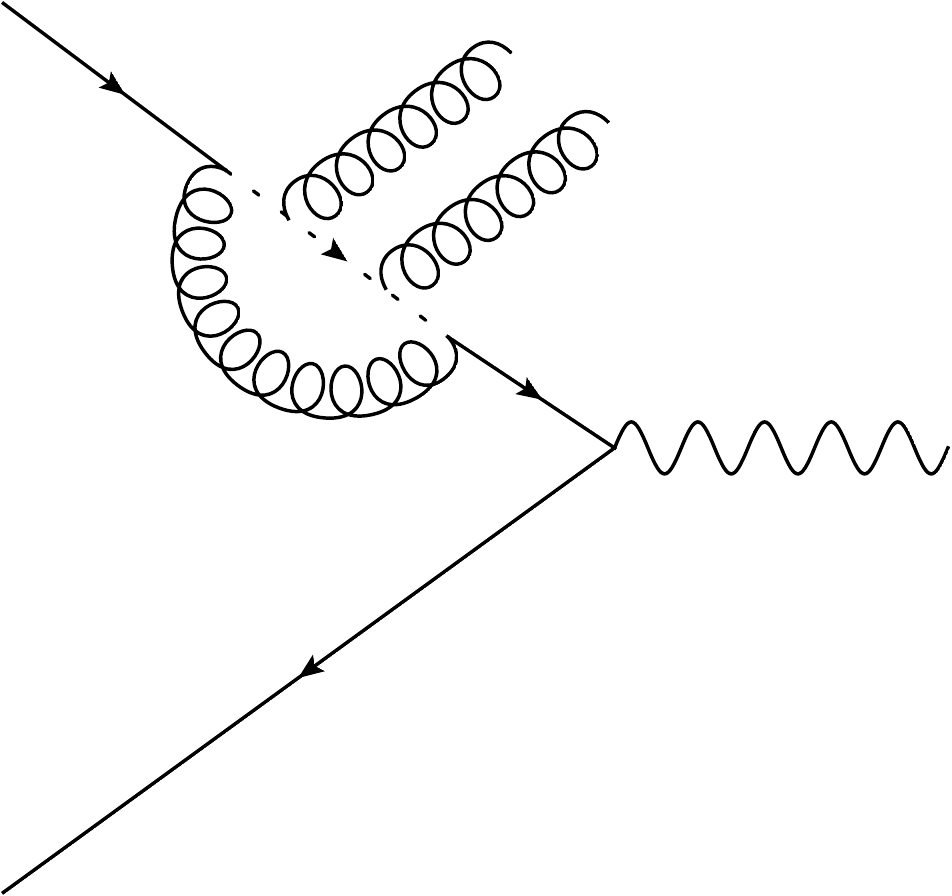}}
\caption{Physical intepretation of the soft region occuring for the
  first time at N$^3$LO: an incoming collinear quark (or antiquark)
  turns into a soft quark by emitting a collinear gluon. The soft
  quark then emits two soft gluons.}
\label{fig:softregion}
\end{center}
\end{figure}

Despite the above cancellation between UV and IR divergences, there
remains the above-mentioned nonzero contribution to the soft region,
which is particularly interesting in that it is new at N$^3$LO in
perturbation theory. To see this, note that we need a virtual gluon in
order to discuss decomposition of the loop momentum. Furthermore, the
new soft region involves the momentum scale $k_1\cdot k_2$, which can
only be formed if there are at least two soft gluons in the final
state. Detailed scrutiny of the region expansion applied to each of
our Feynman diagrams reveals that the sole contribution to the soft
region stems from physical configurations similar to those of
figure~\ref{fig:softregion}. In the example shown, the incoming collinear
quark turns into a soft quark by emitting a collinear gluon, where the
soft quark then emits two soft gluons. As is well-known, soft quarks
are subleading (in the momentum expansion) relative to soft
gluons. Thus, we expect the soft region to contribute (if at all) at
NLP level only. Furthermore, the somewhat complicated structure of
soft and collinear emissions, together with the fact that this region
occurs for the first time at N$^3$LO, suggests that it will be
suppressed by a number of powers of $\epsilon$, so as to give
subleading logs in the final result for the $K$ factor. We will see in
what follows that both of these expectations are borne out. It is also
worth mentioning that a similar soft region was seen in the N$^3$LO
Higgs boson computation of ref.~\cite{Anastasiou:2014lda}, where it
was found to indeed be nonzero. We expect an essentially identical
contribution to appear within the framework of soft collinear
effective theory (SCET).

In summary, careful application of the method of regions to the
process of figure~\ref{fig:loopdiag} reveals the presence of hard,
(anti-)collinear and soft regions. The latter crucially relies on the
presence of a virtual gluon (giving rise to a loop momentum expandable
in regions), as well as two real emissions, to provide the nonzero
scale $k_1\cdot k_2$ associated with the soft region. After expanding
all propagators in each region, all integrals over the loop momentum
$k$ can be carried out analytically. Given that such integrals at
one-loop order are quite standard in the literature, we do not report
intermediate results here. Results for the squared matrix element in
each region can be found in the following section. In order to
cross-check our results, all steps of this calculation (e.g. diagram
calculation, reduction to master integrals, expansion in regions, loop
integration) have been carried out twice, in two completely
independent implementations, and with full agreement.

\subsection{Phase space integration}
\label{sec:PS}

Applying the methods of the previous section, one obtains the
interference term appearing in the integrand of
eq.~(\ref{sigma32r1v}), expanded in regions and integrated over the
loop momentum. The results are compact enough to report here, and it
is first convenient to define the invariants
\begin{align}
t_2&=(p-k_1)^2=-2p\cdot k_1,\notag\\
t_3&=(p-k_2)^2=-2p\cdot k_2,\notag\\
u_2&=(\bar{p}-k_1)^2=-2\bar{p}\cdot k_1,\notag\\
u_3&=(\bar{p}-k_2)^2=-2\bar{p}\cdot k_2,\notag\\
s_{12}&=(k_1+k_2)^2=2k_1\cdot k_2,
\label{tudef}
\end{align}
as well as the combination
\begin{equation}
{\cal M}=\int\frac{d^d k}{(2\pi)^d} 
{\cal A}_{\rm 2r, 1v}(p,\bar{p},k_1,k_2,k)\,
{\cal A}^\dag_{2r}(p,\bar{p},k_1,k_2),
\label{ampsq}
\end{equation}
consisting of the 1-loop double real contribution contracted with the
conjugate tree-level result, integrated over the loop
momentum. Results for the hard region at (next-to) leading power are
then as follows:
\begin{align}
{\cal M}_{\rm hard}^{\rm LP}&={\cal N}\left(\frac{\mu_{\rm
    \overline{MS}}^2}{-s}\right)^{\epsilon} f_1^{\rm H}
\frac{s^3}{t_2\, t_3\, u_2\, u_3};\notag\\ {\cal M}_{\rm hard}^{\rm
  NLP}&={\cal N}\left(\frac{\mu_{\rm
    \overline{MS}}^2}{-s}\right)^{\epsilon}
\frac{s^2(t_2+t_3+u_2+u_3)}{t_2\,t_3\,u_2\,u_3} \left[f_2^{\rm
    H}+\frac{1}{2}\frac{t_2\,u_3+t_3\,u_2-s_{12}\,s}
  {(t_2+t_3)(u_2+u_3)}f_1^{\rm H}\right].
\label{MresultsH}
\end{align}
where
\begin{equation}
{\cal N}=128\pi\,\alpha_s^3(1-\epsilon)
\,C_F^3\, e_q^2\, N_c\,(\mu^2)^{2\epsilon},
\label{normdef}
\end{equation}
and the various functions
$\{f_i^{X}\}$ are defined in appendix~\ref{app:fresults}. Likewise,
the squared matrix element in the collinear region turns out to be
\begin{align}
{\cal M}_{\rm col.}^{\rm LP}&=0;\notag\\
{\cal M}_{\rm col.}^{\rm NLP}&={\cal N}(\mu_{\rm
    \overline{MS}}^2)^\epsilon\frac{s^2}{t_2 t_3 u_2 u_3} \bigg\{
\bigg[u_2 (-t_2)^{-\eps} + u_3(-t_3)^{-\eps} \bigg] f_1^{\rm C} \notag\\
&+\,  \frac{t_3 u_2 + t_2 u_3 - s_{12} s}{t_2 + t_3} 
\bigg[ \bigg((-t_2)^{-\eps} 
-2 (-t_2-t_3)^{-\eps}
+ (-t_3)^{-\eps}\bigg) f^{\rm C}_2 \notag\\ 
&-\, \bigg( \frac{t_2}{t_3} (-t_2)^{-\eps} 
-\frac{(t_2^2 + t_3^2)}{t_2 t_3} (-t_2-t_3)^{-\eps} 
+\frac{t_3}{t_2} (-t_3)^{\eps} \bigg) f_3^{\rm C} 
 \bigg] \bigg\}.
\label{MresultsC}
\end{align}
The anti-collinear region can be straightforwardly obtained through the
exchange $p\leftrightarrow\bar{p}$. Finally, there is the soft region,
which yields
\begin{align}
{\cal M}_{\rm soft}^{\rm LP}&=0;\notag\\
{\cal M}_{\rm soft}^{\rm NLP}&={\cal N}\left(\frac{\mu_{\rm
    \overline{MS}}^2}{-s_{12}}\right)^{\epsilon}
\frac{s^2}{t_2 t_3 u_2 u_3} \notag\\
&\times\bigg\{
\frac{t_3\, f_1^{\rm S}}{t_2 (t_2 + t_3)^2} \bigg[ (s_{12} s - t_2 u_3 - t_3 u_2)
\bigg(t_2 + t_3 - t_3 \, _2 F_1\Big(1,1,1-\eps,\tfrac{t_2}{t_2 + t_3}\Big) \bigg) \bigg] \notag \\ \nn
&+\, \frac{f_2^{\rm S}}{s \, s_{12} (t_2 + t_3)}  
\big[ (t_2 u_3 - t_3 u_2)^2 - s_{12} s(t_2 u_3 + t_3 u_2) \big]  \\  \nn
&+\,\frac{f_3^{\rm S}}{s \, s_{12} t_2 (t_2+t_3)^2} 
\bigg[ s^2_{12} s^2 t_3 (t_2 - t_3) + t_3 (t_2 + t_3)(t_2 u_3 - t_3 u_2)^2 \\ \nn
&+\, s_{12} s t_2 (t_2 + t_3)(t_2 u_3 - 3 t_3 u_2) 
- t_3 \Big(s^2_{12} s^2 (t_2 - t_3) + (t_2 + t_3)(t_2 u_3 - t_3 u_2)^2 \\ \nn
&-\, 2 s_{12} s t_2 (t_2 u_3 + t_3 u_2) 
 \Big)  \, _2F_1\Big(1,1,1-\eps,\tfrac{t_2}{t_2 + t_3}\Big)\bigg]  \\ 
&+\, \{ t_2,t_3 \leftrightarrow u_2, u_3\} 
+ \{ t_2,t_3 \leftrightarrow u_3, u_2\} 
+\{ t_2,u_2 \leftrightarrow t_3, u_3\}  \bigg\}.
\label{MresultsS}
\end{align}
To compute the contribution of
eqs.~(\ref{MresultsH})--(\ref{MresultsS}) to the differential
cross-section or $K$ factor, we must integrate over the
Lorentz-invariant three-body phase space associated with the final
state, as stated in eq.~(\ref{sigma32r1v}). One is free to choose a
particular momentum frame for the phase space
integration. Furthermore, given that each separate term in
eqs.~(\ref{MresultsH})--(\ref{MresultsS}) is Lorentz invariant, we are
free to choose different frames for different types of contribution,
according to convenience. For the hard and collinear regions,
expanding the right-hand side of
eqs.~(\ref{MresultsH})--(\ref{MresultsC}) before substituting into
eq.~(\ref{sigma32r1v}) reveals a series of terms, all containing the
master integral
\begin{align}
I_1(\alpha_1,\alpha_2,\beta_1,\beta_2,\gamma_1,\gamma_2,\delta)
&=\int d\Phi^{(3)}
s_{12}^{\delta}
\,t_2^{-\alpha_1}\,t_3^{-\alpha_2}\,u_2^{-\beta_1}\,u_3^{-\beta_2}
(t_2+t_3)^{-\gamma_1}(u_2+u_3)^{-\gamma_2}. 
\label{I1def}
\end{align}
where $\delta\in\{0,1\}$. For these values of $\delta$, it is possible
to obtain a result for this integral as an expansion the threshold
variable $(1-z)$ for any value of the spacetime dimension $d$, by
decomposing each real gluon momentum $k_i$ in a Sudakov decomposition,
similar to eq.~(\ref{Sudakov}). We spell out this derivation in
appendix~\ref{app:sudint}, and here present the results
\begin{align}
&I_1(\alpha_1,\alpha_2,\beta_1,\beta_2,0,0,0)=
(-1)^{-C}\,2^{-1-2d}\,\pi^{3-2d}\,\Omega^2_{d-2}\,s^{d-3-C}
\frac{(1-z)^{2d-5-C}}{\Gamma(2d-4-C)}\notag\\
&\quad\times\left[\prod_{i=1}^2
\Gamma\left(\frac{d-2}{2}-\alpha_i\right)
\Gamma\left(\frac{d-2}{2}-\beta_i\right)\right]\notag\\
&\quad\times\left[1+(1-z)\left(\frac{\left(\frac{d-2}{2}-\alpha_1\right)
\left(\frac{d-2}{2}-\beta_2\right)+\left(\frac{d-2}{2}-\alpha_2\right)
\left(\frac{d-2}{2}-\beta_1\right)}{2d-4-C}
\right)+{\cal O}[(1-z)^2]\right];\notag\\[15pt]
&I_1(\alpha_1,\alpha_2,\beta_1,\beta_2,\gamma_1,\gamma_2,0)
=(-1)^{-C-\gamma_1-\gamma_2}\,2^{-1-2d}\,\pi^{3-2d}\,\Omega^2_{d-2}\,
s^{d-3-C-\gamma_1-\gamma_2}
\notag\\
&\quad \times\frac{(1-z)^{2d-5-C-\gamma_1-\gamma_2}}{\Gamma(2d-4-C-\gamma_1-\gamma_2)}\left[\prod_{i=1}^2
\Gamma\left(\frac{d-2}{2}-\alpha_i\right)
\Gamma\left(\frac{d-2}{2}-\beta_i\right)\right]\notag\\
&\quad\times\frac{\Gamma(d-2-\alpha_1-\alpha_2-\gamma_1)
\Gamma(d-2-\beta_1-\beta_2-\gamma_2)}{\Gamma(d-2-\alpha_1-\alpha_2)
\Gamma(d-2-\beta_1-\beta_2)(2d-4-C-\gamma_1-\gamma_2)}
\left[1+{\cal O}(1-z)\right];\notag\\[15pt]
&I_1(\alpha_1,\alpha_2,\beta_1,\beta_2,\gamma_1,\gamma_2,1)=
(-1)^{-C-\gamma_1-\gamma_2}\,2^{-1-2d}\,\pi^{3-2d}\,\Omega^2_{d-2}\,
s^{d-2-C-\gamma_1-\gamma_2}\notag\\
&\times
(1-z)^{2d-3-C-\gamma_1-\gamma_2}\left[\prod_{i=1}^2
\Gamma\left(\frac{d-2}{2}-\alpha_i\right)
\Gamma\left(\frac{d-2}{2}-\beta_i\right)\right]\notag\\
&\times \frac{\Gamma(d-1-\alpha_1-\alpha_2-\gamma_1)
\Gamma(d-1-\beta_1-\beta_2-\gamma_2)}{\Gamma(d-1-\alpha_1-\alpha_2)
\Gamma(d-1-\beta_1-\beta_2)\Gamma(2d-2-C-\gamma_1-\gamma_2)}\notag\\
&\times
\left[\left(\frac{d-2}{2}-\alpha_1\right)
\left(\frac{d-2}{2}-\beta_2\right)+
\left(\frac{d-2}{2}-\alpha_2\right)
\left(\frac{d-2}{2}-\beta_1\right)\right]
\left[1+{\cal O}(1-z)\right],
\label{I1results}\end{align}
which are sufficient to integrate eqs.~(\ref{MresultsH})
and~(\ref{MresultsC}) to NLP order in $(1-z)$. Here we have defined
\begin{equation}
C=\sum_{i=1}^2 (\alpha_i+\beta_i),
\label{Cdef}
\end{equation}
as well as the total solid angle in $(d-2)$ spatial dimensions
\begin{equation}
\Omega^{(d-2)}=\frac{2\pi^{\frac{d-2}{2}}}{\Gamma\left(\frac{d-2}{2}\right)}.
\label{O2def}
\end{equation}
For the soft region, we rely on the symmetry of eq. (\ref{MresultsS}) 
under the (combined) exchange of  $p \leftrightarrow \bar{p}$ and 
$k_1 \leftrightarrow k_2$ to reduce the number of distinct terms that 
need to be integrated. There remain two types of terms: (i) those involving
the hypergeometric function ${_2}F_1(1,1;1-\epsilon;t_2/(t_2+t_3))$;
(ii) those without the hypergeometric. Terms of the latter form are
similar to those that occur in the double real emission contribution
to the NNLO Drell-Yan
cross-section~\cite{Hamberg:1990np,Hamberg2002403} (see also
ref.~\cite{Laenen:2010uz} for a recent derivation in the present
notation). To integrate them, one may apply straightforward algebraic
identities such as
\begin{equation}
\frac{1}{t_2(t_2+t_3)}+\frac{1}{t_3(t_2+t_3)}=\frac{1}{t_2\,t_3},\quad
\frac{t_2}{t_3}=\frac{(t_2+t_3)}{t_3}-1
\label{ids}
\end{equation}
(and similarly for $\{u_i\}$) to create a series of terms of the form
of eq.~(\ref{I1def}), with at most one $\alpha_i$ and at most one
$\beta_i$ non-zero. Furthermore, $\delta$ will have a fractional power
that depends on $\epsilon$, due to the presence of the factor
$s_{12}^{-\epsilon}$ in eq.~(\ref{MresultsS}). As described in
refs.~\cite{Hamberg:1990np,Hamberg2002403,Laenen:2010uz}, this
integral can be carried out exactly in the centre of mass frame of the
two final state gluons. We review this derivation in
appendix~\ref{app:softint}.

The most difficult phase space integrals occur in terms of type (i)
above, namely those in the soft region involving a hypergeometric
function. All such terms involve the master integral
\begin{align}
I_2(\alpha_1,\alpha_2,\beta_1,\beta_2,\gamma_1,\gamma_2,\delta)
&=\int d\Phi^{(3)}
s_{12}^{\delta}
\,t_2^{-\alpha_1}\,t_3^{-\alpha_2}\,u_2^{-\beta_1}\,u_3^{-\beta_2}
(t_2+t_3)^{-\gamma_1}(u_2+u_3)^{-\gamma_2}\notag\\
&\quad\times{_2}F_1\left(1,1;1-\epsilon;\frac{t_2}{t_2+t_3}\right),
\label{I2def}
\end{align}
and we note that similar integrals have been carried out for Higgs
boson production in
refs.~\cite{Anastasiou:2013srw,Anastasiou:2015yha}, whose methods
prove very useful for the present study. We proceed as follows. We
first apply identities similar to eq.~(\ref{ids}) to put all terms in
the form where at most one $\alpha_i$ and at most one $\beta_i$ is
nonzero, finding in all cases that $\alpha_2=0$. As we explain in
appendix~\ref{app:softint}, for integrals involving only
$(\alpha_1,\beta_1)$ potentially nonzero, one may use the centre of
mass frame of the outgoing gluons to derive the analytic result (valid
for arbitrary $d$)
\begin{small}
\begin{align}
&I_2(\alpha_1,0,\beta_1,0,\gamma_1,\gamma_2,\delta)=
2^{1-2d}(-1)^{-\alpha_1-\beta_1-\gamma_1-\gamma_2}\,\pi^{1-d}\,
s^{d-3+\delta-\alpha_1-\beta_1-\gamma_1-\gamma_2}\notag\\
&\times
\frac{\Gamma(d-2+\delta-\alpha_1-\gamma_1)\Gamma(d-2+\delta-\alpha_1-\beta_1)
\Gamma(d/2-1+\delta)}
{\Gamma(2d-4+2\delta-\alpha_1-\beta_1
-\gamma_1-\gamma_2)\Gamma(d-2+\delta-\beta_1)\Gamma(d-2+\delta-\alpha_1)
\Gamma(d-2-\alpha_1-\beta_1)}\notag\\
&\times \frac{\Gamma(d/2-1-\beta_1)\Gamma(d/2-1-\alpha_1)
\Gamma(d-2+\delta-\beta_1-\gamma_2)}{\Gamma(d/2-1)}\,
(1-z)^{2d-5+2\delta-\alpha_1-\beta_1-\gamma_1-\gamma_2}\notag\\
&\times{_4}F_3(1,1,d-2+\delta-\alpha_1-\beta_1,d/2-1-\alpha_1;
d-2+\delta-\alpha_1,a+1,d-2-\alpha_1-\beta_1;1)\notag\\
&+\ldots,
\label{I2res1}
\end{align}
\end{small}
where the ellipsis denotes subleading powers of $(1-z)$. This
expression can be easily expanded in $\epsilon$ using the
\texttt{HypExp} package for the hypergeometric
function~\cite{Huber:2005yg,Huber:2007dx}. All necessary values of the
parameters $\{\alpha_i,\beta_i,\gamma_i,\delta\}$ are collected in
appendix~\ref{app:softint}, together with results for each
integral, where for convenience we define
\begin{align}
I_2(\alpha_1,\beta_1,\alpha_2,\beta_2,\gamma_1,\gamma_2,\delta)
&=(4\pi)^{-3+2\epsilon}
\,e^{-2\epsilon\gamma_E}\,
s^{d-3+\delta-C-\gamma_1-\gamma_2}\, (1-z)^{2d-5+2\delta-C-\gamma_1-\gamma_2}
\notag\\
&\quad\times
\hat{I}_2(\alpha_1,\beta_1,\alpha_2,\beta_2,\gamma_1,\gamma_2,\delta)
+\ldots
\label{I2hatdef}
\end{align}
For integrals involving $(\alpha_1,\beta_2)$ non-zero, we were not
able to find any comparable closed form expression. However, they can
be evaluated using Mellin-Barnes techniques, and the ``energies and
angles'' phase space parametrisation described in
refs.~\cite{Anastasiou:2013srw,Anastasiou:2015yha}. We describe this
method in appendix~\ref{app:softint}, but note here that in order to
apply it to integrals involving negative powers of $\gamma_1$ and / or
$\gamma_2$, one must reexpress them in terms of other integrals, some
involving more than two nonzero values of
$(\alpha_1,\alpha_2,\beta_1,\beta_2)$. Results are collected in
appendix~\ref{app:softint}, again using the notation of
eq.~(\ref{I2hatdef}). All aspects of the phase space integration,
including the calculation of all relevant master integrals, have been
carried out twice and completely independently, with full agreement.

\section{Results}
\label{sec:results}

We now have all the necessary ingredients for assembling the
abelian-like terms ($\sim C_F^3$) in the 2-real, 1-virtual
contribution to the $K$ factor of eq.~(\ref{Kdef}), in the $q\bar{q}$
channel up to NLP order~\footnote{As in ref.~\cite{Bonocore:2014wua},
  we will present the unrenormalised $K$ factor.}. We will present
separate results for the hard, (anti-)collinear and soft regions. For
the hard region, one has (in the normalisation of eq.~(\ref{Kdef}))
\begin{align}
&\left.K^{\rm (3), H}_{q\bar{q}}\right|_{C_F^3}=128\notag\\&\quad
\times\left\{\frac{1}{\epsilon^5}\left({\cal D}_0-1\right)
+\frac{1}{\epsilon^4}\left(
-4{\cal D}_1+\frac{3{\cal D}_0}{2}+4L-4\right)
+\frac{1}{\epsilon^3}\left(8{\cal D}_2-6{\cal D}_1+\frac{(8-21\zeta_2)}{2}
{\cal D}_0\right.\right.\notag\\
&\left.\left.\quad-8L^2+16L-\frac{31}{4}+\frac{21}{2}\zeta_2\right)
+\frac{1}{\epsilon^2}\left[-\frac{32{\cal D}_3}{3}
+12{\cal D}_2+(-16+42\zeta_2){\cal D}_1+\left(8-\frac{63}{4}\zeta_2
\right.\right.\right.\notag\\
&\left.\left.\left.\quad-23\zeta_3\phantom{\frac{63}{4}}\hspace{-0.5cm}\right)
{\cal D}_0
+\frac{32}{3}L^3-32L^2+(31-42\zeta_2)L-18+42\zeta_2
+23\zeta_3\right]+\frac{1}{\epsilon}\left[
\frac{32}{3}{\cal D}_4-16{\cal D}_3\right.\right.\notag\\
&\left.\left.\quad+(32-84\zeta_2){\cal D}_2
+(-32+63\zeta_2+92\zeta_3){\cal D}_1+\left(16-42\zeta_2-\frac{69}{2}\zeta_3
+\frac{1017}{16}\zeta_4\right){\cal D}_0-\frac{32}{3}L^4\right.\right.\notag\\
&\left.\left.\quad+\frac{128}{3}L^3+(-62+84\zeta_2)L^2
+\left(72-168\zeta_2-92\zeta_3\right)L
-36+\frac{651}{8}\zeta_2+92\zeta_3-\frac{1017}{16}\zeta_4
\right]\right.\notag\\
&\left.\quad-\frac{128}{15}{\cal D}_5+16{\cal D}_4+\left(-\frac{128}{3}
+112\zeta_2\right){\cal D}_3+(64-126\zeta_2-184\zeta_3){\cal D}_2
+\left(-64+168\zeta_2\phantom{\frac{a}{b}}\right.\right.\notag\\
&\left.\left.\quad+138\zeta_3-\frac{1017}{4}\zeta_4\right){\cal D}_1
+\left(32-84\zeta_2-92\zeta_3+\frac{3051}{32}\zeta_4-\frac{1053}{5}\zeta_5
+\frac{483}{2}\zeta_3\,\zeta_2\right){\cal D}_0+\frac{128}{15}L^5\right.\notag\\
&\left.\quad-\frac{128}{3}L^4
+\left(\frac{248}{3}-112\zeta_2\right)L^3+(-144+336\zeta_2+184\zeta_3)L^2
+\left(144-\frac{651}{2}\zeta_2-368\zeta_3\right.\right.\notag\\
&\left.\left.\quad+\frac{1017}{4}\zeta_4\right)L
\right\}
,
\label{Khard}
\end{align}
where (given the focus of our study) we report only enhanced
(non-constant) terms in the finite part, and we have made the
conventional choice
\begin{equation}
\mu^2_{\rm \overline{MS}}=4\pi e^{-\gamma_E}\mu^2=Q^2
\end{equation}
for the dimensional regularisation scale in the $\overline{\rm MS}$
scheme. NLP terms will be sensitive to this choice, given that the K
factor contains the dimensional combination
\begin{displaymath}
\left(\frac{\bar{\mu}^2}{s}\right)^\epsilon\rightarrow 
\left(\frac{Q^2}{s}\right)^\epsilon=z^\epsilon. 
\end{displaymath}
Note that we have identified
\begin{displaymath}
\frac{\log^{n}(1-z)}{1-z}\rightarrow 
\left[\frac{\log^{n}{(1-z)}}{1-z}\right]_+\equiv{\cal D}_n
\end{displaymath}
everywhere, i.e. we have neglected the delta function contribution
that arises from rewriting LP terms in terms of plus
distributions. The delta function terms mix with virtual corrections
not included here, and thus are not worth reporting. For the collinear
region, we find
\begin{align}
&\left.K^{\rm (3), C}_{q\bar{q}}\right|_{C_F^3}=
32\left\{-\frac{1}{\epsilon^4}+\frac{1}{\epsilon^3}\left(5L-\frac54\right)
+\frac{1}{\epsilon^2}\left(-\frac32-\frac{25}{2}L^2+\frac{25}{4}L
+\frac{21}{2}\zeta_2\right)\right.\notag\\
&\left.\quad+\frac{1}{\epsilon}\left[\frac{125L^3}{6}-\frac{125L^2}{8}
+\left(\frac{15}{2}-\frac{105\zeta_2}{2}\right)L
-2+\frac{105}{8}\zeta_2+41\zeta_3\right]
-\frac{625}{24}L^4+\frac{625}{24}L^3\right.\notag\\
&\left.\quad+\left(-\frac{75}{4}+\frac{525\zeta_2}{4}
\right)L^2+\left(10-\frac{525}{8}\zeta_2-205\zeta_3\right)L
\right\},
\label{Kcoll}
\end{align}
where the anti-collinear region gives an identical
contribution. Finally, we have the soft region, whose contribution is
\begin{align}
&\left.K^{\rm (3), S}_{q\bar{q}}\right|_{C_F^3}=
32\left\{\frac{1}{\epsilon}\left(\frac23\zeta_2+\frac13\zeta_3\right)
-(4\zeta_2+2\zeta_3)L\right\}.
\label{Ksoft}
\end{align}
The total result for the (unrenormalised) $K$ factor up to NLP order in
the threshold expansion can be obtained from the above results through
the combination
\begin{equation}
\left.K^{(3)}_{q\bar{q}}\right|_{C_F^3}=\Big[K^{(3), H}+2
K^{(3), C}+K^{(3), S}\Big]_{C_F^3}.
\label{Ktot}
\end{equation}
Equations~(\ref{Khard}, \ref{Kcoll}, \ref{Ksoft}) constitute the main
results of this paper. As discussed above, our main motivation for
presenting them is as a prerequisite for formulating and testing
general prescriptions for classifying (and potentially resumming) NLP
threshold corrections in arbitrary processes. Although a full study in
this regard is beyond the scope of this paper, it is worthwhile to
make a few remarks regarding the implications of our results.

Following a detailed analysis of the 1-real, 1-virtual $K$ factor in the
$q\bar{q}$ channel~\cite{Bonocore:2014wua},
refs.~\cite{Bonocore:2015esa} considered a general amplitude with $N$
hard particles, which is then dressed by an extra gluon emission. A
process-independent factorisation formula was presented, building on
the earlier work of ref.~\cite{DelDuca:1990gz}, which captured all
abelian-like contributions to the amplitude up to NLP order in the
threshold expansion, in the absence of final state jets. This formula
was generalised to include all fully non-abelian contributions in
ref.~\cite{Bonocore:2016awd}, and extends the well-known
soft-collinear factorisation formula for LP threshold effects (see
e.g.~\cite{Dixon:2008gr}). It includes a number of universal functions
describing soft and collinear behaviour, whose operator definitions
involve (generalised) Wilson
lines~\cite{Laenen:2008gt,White:2011yy}. A new function occuring at
NLP level is the so-called {\it jet emission function}, first
introduced in ref.~\cite{DelDuca:1990gz}. As its name suggests, it
describes the dressing of a jet function (collecting virtual collinear
effects) with an additional radiative gluon. A fully non-abelian
operator definition for this quantity has been proposed for
(anti-)quark jets in ref.~\cite{Bonocore:2016awd}, and calculated at
one-loop order. A similar calculation is in progress for gluons, which
would have immediate applications in e.g. Higgs boson production via
gluon-gluon fusion.

In processes containing two or more additional gluons, an open
question is whether the functions appearing in the one-emission case
are sufficient to capture all physics up to NLP order in the threshold
expansion, or whether new functions should appear. For example, one
may consider generalising the jet emission function to a family of
quantities representing the dressing of a nonradiative jet with
arbitrary numbers of additional gluons. For resummation of NLP effects
to be possible, it should ideally be the case that these higher
multiplicity jet emission functions are related by an iterative
property to those with lower numbers of emissions (for a preliminary
discussion in a purely abelian context, see
ref.~\cite{DelDuca:1990gz}). Or, this may be possible only up to a
given subleading logarithmic order.

At NNLO in Drell-Yan production, it was already noticed that, perhaps
unsurprisingly, the (anti-)collinear region in the method of regions
maps straightforwardly to the contribution of the jet emission
functions associated with the incoming (anti-)quark legs in the
factorisation approach. Furthermore, this contribution started only at
next-to-leading-logarithmic (NLL) order, and at next-to-leading power
(NLP) in the threshold variable. In the present calculation, we also
see that the (anti-)collinear regions start only at ${\cal
  O}(\epsilon^{-4})$ rather than ${\cal O}(\epsilon^{-5})$. Thus,
again we find that collinear effects are NLP, and give only subleading
(NLL) threshold logarithms. Indeed, the only source of leading LP or
NLP effects is the hard region, as can be clearly seen in
eq.~(\ref{Khard}). This observation will certainly be a useful guide
when examining the extent to which (multiple) jet emission functions
are relevant at higher orders in perturbation theory. Furthermore,
there is much existing evidence (most notably in
ref.~\cite{Moch:2009hr}) that the highest power of the NLP log
exponentiates in Drell-Yan. The observation that collinear effects do
not affect this log at N$^3$LO provides a significant hint regarding
how to formally prove this property.

The new soft region at this order depends crucially on the presence of
two gluons, and so would seem to be a correction to factorisation
formulae of the type presented in
refs.~\cite{Bonocore:2015esa,Bonocore:2016awd}, in that it cannot be
composed iteratively from lower-order information. However, it is
worthwhile to note that the soft region itself is heavily suppressed
in the $\epsilon$ expansion, so that it only contributes logarithmic
terms at N$^4$LL level. If this behaviour persists at higher orders,
such a region is unlikely to trouble realistic efforts to resum NLP
effects, but it should of course be fully understood, as it will be
present in the exact Drell-Yan $K$ factor at higher orders. 

Further insights into the iterative structure of our results can be
obtained by examining the squared matrix elements before integration
over the final state phase space, but after the integration over the
loop momentum of the virtual gluon. In the case of the hard region
(eq.~(\ref{MresultsH})), we find that the coefficient $f_1^{\rm H}$
matches the similar function found in the one-loop quark form factor,
such that the leading power term agrees with what one obtains from
applying the well-known eikonal Feynman rules to the non-radiative
one-loop Drell-Yan process. At NLP, we noted that the second
coefficient $f_2^{\rm H}$ already appears in the 1-real, 1-virtual
contributions at NNLO. Thus, there is strong evidence that the hard
region can indeed be understood using the existing tools of
refs.~\cite{Laenen:2008gt,Laenen:2010uz,Bonocore:2015esa,Bonocore:2016awd}. In
the collinear region, we find that the function $f_1^{\rm C}$ in the
first line of eq.~(\ref{MresultsC}) occurs already at NNLO, such that
this contribution factorises into a one-loop jet emission on the quark
leg, dressed by a tree-level emission from the anti-quark (and vice
versa for the anti-collinear region). The remaining collinear
contributions, involving the additional coefficients $f_{2,3}^{\rm
  C}$, lack such a straightforward interpretation, leaving open the
possibility that one must consider a separate jet emission function
for pairs of gluons. Finally, as discussed already above, the soft
region is not expected to be iteratively obtainable from lower order
information.

\section{Conclusion}
\label{sec:conclude}

In this paper, we examine abelian-like contributions to Drell-Yan
production in the $q\bar{q}$ channel at N$^3$LO, namely those with the
colour structure $\propto C_F^3$. We have classified all
logarithmically enhanced contributions near threshold when one gluon
is virtual, and the other two real, up to next-to-leading power (NLP)
in the threshold variable $(1-z)$. Our motivation is to work towards a
systematic classification of NLP threshold effects, building on
e.g. the factorisation formulae of
refs.~\cite{Bonocore:2015esa,Bonocore:2016awd} (see
refs.~\cite{Larkoski:2014bxa,Kolodrubetz:2016uim,
  Moult:2016fqy,Boughezal:2016zws,Moult:2017rpl,Chang:2017atu,Feige:2017zci}
for similar work within the context of effective field theory). To
this end, we present results for the unrenormalised $K$ factor, using
the method of regions~\cite{Beneke:1997zp,Pak:2010pt,Jantzen:2011nz}
to separate contributions according to whether the virtual gluon is
hard, soft or collinear with one of the incoming particles. Our hope
is that this provides a great deal of useful information for
elucidating the general structure of NLP effects, similar to how
previous methods of region analyses at NNLO~\cite{Bonocore:2014wua}
directly informed the construction of factorisation formulae valid to
subleading order in the threshold expansion.

There are a number of noteworthy features in our result. Firstly,
there is a nonzero soft region that appears for the first time at
N$^3$LO, and which we find persists upon integration over the final
state phase space. The presence of such a contribution requires at
least one virtual gluon and two real gluons, and thus does not appear
to be iteratively relatable to lower order information. A similar
region was found to be nonzero in the recent (and closely related)
calculation of Higgs boson production via gluon-gluon
fusion~\cite{Anastasiou:2015yha}, whose methods prove very useful for
the present analysis. The overall contribution of this region to the
Drell-Yan $K$ factor is highly subleading, in that it contributes with a
single pole in the dimensional regularisation parameter $\epsilon$ at
${\cal O}(\alpha_s^3)$, corresponding to a N$^4$LL NLP logarithm in
the finite part of the $K$ factor. It would be interesting to see what
effect such a region has at higher orders in perturbation theory, and
indeed whether it has a straightforward counterpart in SCET.

Unlike the hard region, the collinear region does not contribute to
the leading NLP logarithm, suggesting that collinear effects are not
relevant to the potential resummation of the highest power of NLP logs
to all orders in perturbation theory. Both the hard and collinear
regions in our analysis show signs of an iterative structure, whereby
parts of the results can be obtained from lower order
information. These observations will prove highly useful in
generalising factorisation formulae for NLP effects to higher orders
in perturbation theory.

There are a number of directions for further work. Immediately related
to the present study would be the calculation of threshold
contributions in the triple real emission contributions to Drell-Yan
production at N$^3$LO, or in the double-virtual, single real
channel. Furthermore, one can generalise the calculation to include
all possible colour structures, involving fully non-abelian
corrections. Finally, the implications of our results for developing a
fully systematic classification of NLP threshold effects in arbitrary
scattering processes will be the subject of much further study.

\section*{Acknowledgments}

We thank Domenico Bonocore, Eric Laenen and Lorenzo Magnea for
discussion and collaboration on related topics, and for comments on
the manuscript. We are especially grateful to Claude Duhr and \"{O}mer
G\"{u}rdo\u{g}an for detailed advice and, in the latter case,
Mathematica code relating to Mellin-Barnes integration. We thank
Andreas von Manteuffel for communications related to the computer
program \texttt{Reduze}. In addition, we thank Martin Beneke and Einan
Gardi for useful conversations. This work was supported by the Dutch
National Organization for Scientific Research (NWO), the D-ITP
consortium, a program of NWO funded by the Dutch Ministry of
Education, Culture and Science (OCW), and by the UK Science and
Technology Facilities Council (STFC).


\appendix

\section{Coefficients entering the matrix element}
\label{app:fresults}

In this appendix, we collect results for the various coefficients
appearing in eqs.~(\ref{MresultsH})--(\ref{MresultsS}). Starting with
the hard region, we have
\begin{align}
f_1^{\rm H}&=-\frac{2}{\epsilon^2}-\frac{3}{\epsilon}-8+\zeta_2
+\epsilon\left(-16+\frac{3\zeta_2}{2}+\frac{14\zeta_3}{3}\right)
+\epsilon^2\left(-32+4\zeta_2+7\zeta_3+\frac{47\zeta_4}{8}\right)\notag\\
&+\epsilon^3\left(-64+8\zeta_2+\frac{56\zeta_3}{3}+\frac{141\zeta_4}{16}
+\frac{62\zeta_5}{5}-\frac{7}{3}\zeta_3\zeta_2\right)
+\epsilon^4\left(-128+16\zeta_2+\frac{112\zeta_3}{3}\right.\notag\\
&\left.+\frac{47\zeta_4}{2}
+\frac{93\zeta_5}{5}-\frac{7\zeta_2\zeta_3}{2}+\frac{949\zeta_6}{64}
-\frac{49\zeta_3^2}{9}\right)+\ord(\eps^5);\notag\\
f_2^{\rm H}&=(1-\epsilon)f_1^{\rm H}.
\label{fHresults}
\end{align}
The coefficients for the (anti-)collinear regions are
\begin{align}
f_1^{\rm C}&= - \frac{2}{\eps}-\frac{5}{2} 
+ \eps \big(-3 + \zeta_2\big) 
+ \eps^2 \bigg(-4 + \frac{5 \zeta_2}{4} + \frac{14 \zeta_3}{3} \bigg) 
+ \eps^3 \bigg(-6 + \frac{3 \zeta_2}{2} + \frac{35 \zeta_3}{6} 
+ \frac{47 \zeta_4}{8} \bigg) \notag\\
&+\, \eps^4 \bigg(-10 + 2 \zeta_2 + 7 \zeta_3  
+ \frac{235 \zeta_4}{32} + \frac{62 \zeta_5}{5}- \frac{7\zeta_2 \zeta_3}{3} \bigg) +\ord(\eps^5);\notag\\
f_2^{\rm C}&=- \frac{1}{4 \eps} + \frac{1}{8} 
+ \eps \bigg(\frac{3}{4} + \frac{\zeta_2}{8}\bigg) 
+ \eps^2 \bigg(2 - \frac{\zeta_2}{16} + \frac{7 \zeta_3}{12} \bigg) 
+ \eps^3 \bigg(\frac{9}{2} - \frac{3 \zeta_2}{8} - \frac{7 \zeta_3}{24} 
+ \frac{47 \zeta_4}{64} \bigg) \notag\\
&+\, \eps^4 \bigg(\frac{19}{2} - \zeta_2 - \frac{7 \zeta_3}{4} 
  - \frac{47 \zeta_4}{128} 
+ \frac{31 \zeta_5}{20}- \frac{7\zeta_2 \zeta_3}{24} \bigg) + \ord(\eps^5)
;\notag\\
f_3^{\rm C}&=\frac{1}{4\epsilon^2}-\frac{1}{8\epsilon}-\frac34
-\frac{\zeta_2}{8}+\epsilon\left(-2+\frac{\zeta_2}{16}-\frac{7\zeta_3}{12}
\right)+\epsilon^2\left(-\frac92+\frac{3\zeta_2}{8}+\frac{7\zeta_3}{24}
-\frac{47\zeta_4}{64}\right)\notag\\
&+\epsilon^3\left(-\frac{19}{2}+\zeta_2+\frac{7\zeta_3}{4}
+\frac{47\zeta_4}{128}-\frac{31\zeta_5}{20}+\frac{7}{24}\zeta_2\,\zeta_3
\right)+\epsilon^4\left(-\frac{39}{2}+\frac{9\zeta_2}{4}
+\frac{14\zeta_3}{3}+\frac{141\zeta_4}{64}\right.\notag\\
&\left.+\frac{31\zeta_5}{40}
-\frac{7\zeta_2\,\zeta_3}{48}
-\frac{949\zeta_6}{512}+\frac{49\zeta_3^2}{72}\right)+\ord(\eps^5).
\label{fCresults}
\end{align}
For the soft region, we have
\begin{align}
f_1^{\rm S}&=\frac{1}{4 \eps^2} + \frac{1}{4 \eps} 
+ \frac{1}{2} - \frac{\zeta_2}{8} 
+ \eps \bigg(1 - \frac{\zeta_2}{8} - \frac{7 \zeta_3}{12} \bigg) 
+ \eps^2 \bigg(2 - \frac{\zeta_2}{4} - \frac{7 \zeta_3}{12} 
- \frac{47 \zeta_4}{64} \bigg) \notag\\ \nn
&+\, \eps^3 \bigg(4 - \frac{\zeta_2}{2} 
- \frac{7 \zeta_3}{6}  - \frac{47 \zeta_4}{64} 
- \frac{31 \zeta_5}{20}+ \frac{7\zeta_2 \zeta_3}{24} \bigg) \\ \nn 
&+\,  \eps^4 \bigg(8 - \zeta_2 - \frac{7 \zeta_3}{3}  - \frac{47 \zeta_4}{32} 
- \frac{31 \zeta_5}{20} + \frac{7 \zeta_2 \zeta_3}{24} 
- \frac{949 \zeta_6}{512} + \frac{49 \zeta_3^2}{72}\bigg) 
+ \ord(\eps^5);\notag\\
f_2^{\rm S}&=\frac{1}{4 \eps} + \frac{1}{2}
+ \eps \bigg(1 - \frac{\zeta_2}{8}\bigg) 
+ \eps^2 \bigg(2 - \frac{\zeta_2}{4} - \frac{7 \zeta_3}{12} \bigg)
+ \eps^3 \bigg(4 - \frac{\zeta_2}{2} - \frac{7 \zeta_3}{6} 
 - \frac{47 \zeta_4}{64}\bigg)\notag \\ \nn
&+\, \eps^4 \bigg(8 -\zeta_2 
- \frac{7 \zeta_3}{3} - \frac{47 \zeta_4}{32} - \frac{31 \zeta_5}{20} + \frac{7 \zeta_2 \zeta_3}{24} \bigg) 
+ \ord(\eps^5);\notag\\
f_3^{\rm S}&=
\frac{1}{4 \eps} + \frac{1}{4} 
+ \eps \bigg( \frac{1}{2} - \frac{\zeta_2}{8} \bigg) 
+ \eps^2 \bigg(1 - \frac{\zeta_2}{8} - \frac{7 \zeta_3}{12} \bigg) 
+ \eps^3 \bigg(2 - \frac{\zeta_2}{4} - \frac{7 \zeta_3}{12} 
- \frac{47 \zeta_4}{64} \bigg)\notag \\
&+\, \eps^4 \bigg(4 - \frac{\zeta_2}{2} 
- \frac{7 \zeta_3}{6} - 
\frac{47 \zeta_4}{64} - \frac{31 \zeta_5}{20} + \frac{7 \zeta_2 \zeta_3}{24} \bigg)
+ \ord(\eps^5).
\label{fSresults}
\end{align}

\section{Phase space integrals in the hard and (anti-)collinear regions}
\label{app:sudint}

In this appendix, we spell out the derivation of eq.(\ref{I1results}),
using the Sudakov decomposition of
eqs.(\ref{Sudakov}-\ref{kperpdots}). Furthermore, we define the
quantities $k_{i+}=n_- \cdot k_i$ and $k_{i-}=n_+ \cdot k_i$, using a
slightly different convention to the Sudakov decomposition of the loop
momentum in section \ref{sec:regions}, so as to make factors of 2 more
convenient in the following. The 3-body phase space in $d$ dimensions
is given by
\begin{align}\int d\Phi^{(3)}=\,(2 \pi)^d \int \frac{d^dq}{(2 \pi)^{d-1}}
&\left( \prod_{i=1}^2  \int \frac{d^dk_i}{(2 \pi)^{d-1}} \delta_+(k_i^2) \right) \notag \\ &\times\, \delta_+(q^2-Q^2)
\delta^{(d)}\left(q+\sum_{j=1}^2 k_j- (p+\bar{p})\right)
\label{eqn:ps}
\end{align}
where 
\begin{equation}
\delta_+(k^2)=\theta(k^0)\delta(k^2)
\label{eqn:dplus}
\end{equation}
and $\theta$ is the Heaviside function
\begin{equation}
\theta(k^0)=\begin{cases}
    k^0 & \text{if }  k^0 > 0 \\
    0 & \text{otherwise}.
  \end{cases}
  \label{eqn:heavi}
\end{equation}
We may carry out the integral over the photon momentum $q$ using the
delta function in eq.~(\ref{eqn:ps}), obtaining
\begin{align}
\int d\Phi^{(3)} &= (2 \pi)^{3-2d} 
\left( \prod_{i=1}^2 \int d^dk_i
\delta_+(k_i^2) 
\right)
\delta\left[\left((p+\bar{p})-\sum_{j=1}^2 k_j\right)^2-Q^2\right]\notag\\
&=(2\pi)^{3-2d}\left[\prod_{i=1}^2\frac12 \int d k_{i+}\,dk_{i-}
\,d^{d-2} k_{i\perp}\,\delta_+(k_i^2) 
\right]\notag\\
&\quad\times\delta[(1-z)s-2(k_1+k_2)\cdot (p+\bar{p})
+2k_1\cdot k_2],
\label{eqn:ps1}
\end{align} 
where in the second line we have used eq.~(\ref{zdef}). The delta
function in the last line can be expressed as a Fourier transform:
\begin{equation}
\delta[(1-z)s-2(k_1+k_2)\cdot (p+\bar{p})
+2k_1\cdot k_2]
=\frac{1}{s}\int^{\infty}_{-\infty}\frac{d\omega}{2\pi}e^{i\omega (1-z)}
e^{\frac{-2i\omega}{s} (k_1 \cdot p+k_2 \cdot p + k_1 \cdot \bar{p}+ k_2 \cdot \bar{p})}
e^{\frac{2i\omega}{s}  k_1 \cdot k_2},
\end{equation} 
where we can Taylor expand the exponential in $k_1 \cdot k_2$,
given that higher order terms will be suppressed by powers of $1-z$:
\begin{equation}
e^{\frac{2i\omega}{s}  k_1 \cdot k_2} = 1+\frac{2 i \omega}{s}k_1 \cdot k_2
+{\cal O}(k_i^4). 
\label{eqn:taylor}
\end{equation} 
Putting things together, the phase space becomes
\begin{align}
\int d\Phi^{(3)} &= \frac{(2 \pi)^{3-2d}}{2^2s}
\prod_{i=1}^2 \int^{\infty}_0 dk_{i+}\int^{\infty}_0 dk_{i-} 
\int^{\infty}_{-\infty}d^{d-2}k_{i\perp}
\delta(k_{i+}k_{i-}-|k_{i\perp}|^2) \notag \\
&\times
\int^{i\infty}_{-i\infty}\frac{d\tilde{\omega}}{2 \pi i}e^{\tilde{\omega} (1-z)}
e^{\frac{-\tilde{\omega}}{\sqrt{s}} \sum_{j=1}^{2} (k_{j+} + k_{j-})}  \notag \\
&
\times\left[1+\frac{2\tilde{\omega}}{s} \left(\frac{1}{2}
(k_{1+}k_{2-}+k_{1-}k_{2+})-k_{1\perp} \cdot k_{2\perp}\right)\right],
\label{eqn:HardColPS}
\end{align}
where we have transformed $\tilde{\omega}=i\omega$. We can now use
this result to carry out the integral of eq.~(\ref{I1def}) for the
two special cases of $\delta\in\{0,1\}$.

For $\delta=0$, we may note that the integrand of eq.(\ref{I1def}) has
no transverse momentum dependence, such that the linear term
$k_{1\perp} \cdot k_{2\perp}$ in eq.~(\ref{eqn:HardColPS}) leads to an
odd integrand, and can be neglected. Using polar coordinates for the
$k_{i\perp}$ integrals, one may use the onshell delta functions to
eliminate the integral over $|k_{i\perp}|$, such that
eq.~(\ref{I1def}) becomes
\begin{align}
I_1(\alpha_1,\alpha_2,\beta_1,\beta_2,\gamma_1,\gamma_2,0)
&=(-1)^{(C+\gamma_1+\gamma_2)}\frac{(2 \pi)^{3-2d}}{2^{4}} s^{-1-\frac{1}{2}(C+\gamma_1+\gamma_2)}\Omega^2_{d-2}
\int^{i\infty}_{-i\infty}\frac{d\tilde{\omega}}{2\pi i}e^{\tilde{\omega} (1-z)}
  \notag \\
& \times
\int^{\infty}_0 dk_{1+} e^{\frac{-\tilde{\omega}}{\sqrt{s}} k_{1+}}k_{1+}^{\frac{d-4}{2}-\beta_1}
\int^{\infty}_0 dk_{2+} e^{\frac{-\tilde{\omega}}{\sqrt{s}} k_{2+}}k_{2+}^{\frac{d-4}{2}-\beta_2} 
\left(\frac{1}{k_{1+}+k_{2+}} \right)^{\gamma_2} \notag \\
& \times 
\int^{\infty}_0 dk_{1-} e^{\frac{-\tilde{\omega}}{\sqrt{s}} k_{1-}}k_{1-}^{\frac{d-4}{2}-\alpha_1} 
\int^{\infty}_0 dk_{2-} e^{\frac{-\tilde{\omega}}{\sqrt{s}} k_{2-}}k_{2-}^{\frac{d-4}{2}-\alpha_2}
\left(\frac{1}{k_{1-}+k_{2-}} \right)^{\gamma_1} \notag \\
& \times 
\left( 1+\frac{\tilde{\omega}}{s}
(k_{1+}k_{2-}+k_{1-}k_{2+})\right).
\end{align}
After a variable change
$\tilde{k}_{i\pm}=\frac{\tilde{\omega}}{\sqrt{s}}k_{i\pm}$, we may
recognize the inverse Laplace transform
\begin{equation}
\int_{-i\infty}^{i\infty}\frac{d\tilde{\omega}}{2\pi i}e^{\tilde{\omega}(1-z)} \left(\frac{1}{\tilde{\omega}}\right)^{m} 
=\frac{(1-z)^{m+1}}{\Gamma(m)}. \label{eqn:invlap}
\end{equation}
The integrals over $\tilde{k}_{i\pm}$ will be of the form:
\begin{align*}
\int^\infty_0 d\tilde{k}_{2\pm} \ e^{-\tilde{k}_{2\pm}}\tilde{k}_{2\pm}^n
\int_0^\infty d\tilde{k}_{1\pm} \ e^{-\tilde{k}_{1\pm}}\tilde{k}_{1\pm}^m 
\left(\frac{1}{\tilde{k}_{1\pm}+\tilde{k}_{2\pm}}\right)^l,
\end{align*}
for which the variable transformation
\begin{align*}
\tilde{k}_{1\pm}&=\Lambda w \ ; \ \ \ \ \tilde{k}_{2\pm}=\Lambda (1-w) 
\end{align*}
yields
\begin{align}
\int^1_0 dw \ w^m(1-w)^n
\int_0^\infty d\Lambda \ e^{-\Lambda} \Lambda^{m+n+1-l}  =
\frac{\Gamma(m+1)\Gamma(n+1)}{\Gamma(m+n+2)}\Gamma(m+n-l+2).
\end{align}
Substituting these results, we obtain eq.~(\ref{I1results}) as
required. 

The integral of eq.~(\ref{I1def}) with $\delta=1$ appears only at NLP
level, such that we may entirely neglect the term $k_1\cdot k_2$ in
eq.~(\ref{eqn:taylor}), as it will lead to terms suppressed by further
powers of $(1-z)$. Carrying out similar steps to the $\delta=0$ case,
one again finds eq.~(\ref{I1results}).

\section{Phase space integrals in the soft region}
\label{app:softint}

In this appendix, we describe various integrals (of increasing
complexity) that occur when integrating the squared matrix element in
the soft region (eq.~(\ref{I2res1})) over the final state phase space. 

\subsection{Integrands with no hypergeometric function}
\label{sec:nohyp}

First, we need integrals of the form of eq.~(\ref{I1def}), in which at
most one parameter $\{\alpha_i\}$ and at most one parameter
$\{\beta_i\}$. The Sudakov decomposition of
appendix~(\ref{app:sudint}) turns out not to be helpful here, due to
the fractional power of $\delta$. Instead, one may simplify the
calculation by working in the centre of mass frame of the two outgoing
gluons~\cite{Hamberg:1990np,Hamberg2002403,Laenen:2010uz}. In this
frame, one writes
\begin{align}
k_1&=\frac{\sqrt{s_{12}}}{2}(1,0,\ldots,\sin\theta_2\sin\theta_1,\cos\theta_2
\sin\theta_1,\cos\theta_1),\notag\\
k_2&=\frac{\sqrt{s_{12}}}{2}(1,0,\ldots,-\sin\theta_2\sin\theta_1,-\cos\theta_2
\sin\theta_1,-\cos\theta_1),\notag\\
p&=\frac{(s-\tilde{t})}{2\sqrt{s_{12}}}(1,0,\ldots,0,1),\notag\\
Q&=\left(\frac{s-Q^2-s_{12}}{2\sqrt{s_{12}}},0,\ldots,0,|\vec{q}|\sin\psi,
|\vec{q}|\cos\psi\right),\notag\\
\bar{p}&=\left(\frac{\tilde{t}+s_{12}-Q^2}{2\sqrt{s_{12}}},0,\ldots,0,|\vec{q}|\sin\psi,
|\vec{q}|\cos\psi-\frac{(s-\tilde{t})}{2\sqrt{s_{12}}}\right),
\label{momparam}
\end{align}
where
\begin{align}
\tilde{t}&\equiv 2p\cdot Q=(p+Q)^2-Q^2,\notag\\
\tilde{u}&\equiv 2\bar{p}\cdot Q=(\bar{p}+Q)^2-Q^2,\notag\\
s_{12}&\equiv2k_1\cdot k_2=s-\tilde{t}-\tilde{u}+Q^2,\notag\\
\cos\psi&=\frac{(s-Q^2)(\tilde{u}-Q^2)-s_{12}(\tilde{t}+Q^2)}
{(s-\tilde{t})\sqrt{\lambda(s,Q^2,s_{12})}},\notag\\
|\vec{q}|&=\frac{\sqrt{\lambda(s,Q^2,s_{12})}}{2\sqrt{s_{12}}},
\label{mandies}
\end{align}
and $\lambda$ is the K\"{a}llen function
$\lambda(a,b,c)=a^2+b^2+c^2-2ab-2ac-2bc$. The Mandelstam invariants
$\tilde{t}$ and $\tilde{u}$ can in turn be expressed as functions of
the photon energy fraction $z=Q^2/s$ and of two further variables
$0<x<1$ and $0<y<1$, such that
\begin{align}
\tilde{u}&=s\left[1-y(1-z)\right]\notag\\
\tilde{t}&=s\left[z+y(1-z)-\frac{y(1-y)x(1-z)^2}{1-y(1-z)}\right],
\label{xydef}
\end{align}
where $(1-z)$ is the threshold variable. The 3-body phase space in $d$
dimensions now takes the form
\begin{align}
\int d\Phi^{(3)}&=\frac{1}{(4\pi)^d}\frac{s^{d-3}}{\Gamma(d-3)}(1-z)^{2d-5}
\int_0^\pi d\theta_1\int_0^\pi d\theta_2 (\sin\theta_1)^{d-3}
(\sin\theta_2)^{d-4}\notag\\
&\quad\times\int_0^1dy\int_0^1dx [y(1-y)]^{d-3}[x(1-x)]^{d/2-2}
[1-y(1-z)]^{1-d/2}.
\label{dPS3}
\end{align}
In terms of the above definitions, one finds
\begin{align}
p\cdot k_1&=\frac{s-\tilde{t}}{4}(1-\cos\theta_1)\notag\\
p\cdot k_2&=\frac{s-\tilde{t}}{4}(1+\cos\theta_1)\notag\\
\bar{p}\cdot k_1&=A-B\cos\theta_1-C\sin\theta_1\cos\theta_2\notag\\
\bar{p}\cdot k_2&=A+B\cos\theta_1+C\sin\theta_1\cos\theta_2,
\label{dots}
\end{align}
where
\begin{align}
A&=\frac{\tilde{t}+s_{12}-Q^2}{4},\notag\\
B&=\frac{\sqrt{s_{12}}}{2}|\vec{q}|\cos\psi-\frac{(s-\tilde{t})}{4},\notag\\
C&=\frac{\sqrt{s_{12}}}{2}|\vec{q}|\sin\psi.
\label{ABCdef}
\end{align}
These quantities satisfy the relation
\begin{equation}
A^2=B^2+C^2,
\label{ABCrel}
\end{equation}
such that upon defining 
\begin{equation}
\cos\chi=\frac{B}{A},
\label{coschidef}
\end{equation}
and using the above definitions, the angular integral may be carried
out using the result~\cite{Hamberg:1990np} (first derived
in~\cite{vanNeerven:1985xr})
\begin{align}
&\int_0^\pi d\theta_1\int_0^\pi d\theta_2
\frac{\sin^{d-3}\theta_1 \sin^{d-4}\theta_2}{(1-\cos\theta_1)^p
(1-\cos\chi\cos\theta_1-\sin\chi\sin\theta_1\cos\theta_2)^q}\notag\\
&=2^{1-p-q}\pi\frac{\Gamma(\frac{d}{2}-1-q)\Gamma(\frac{d}{2}-1-p)\Gamma(d-3)}
{\Gamma(d-2-p-q)\Gamma^2(\frac{d}{2}-1)} {_2F}_1\left[p,q;\frac{d}{2}-1;
\cos^2\left(\frac{\chi}{2}\right)\right].
\label{Ipq}
\end{align}
At this stage, one must carry out the integrals over the variables $x$
and $y$ appearing in eq.~(\ref{dPS3}). These can all be carried out in
terms of beta functions, or using the identity
\begin{equation}
\int_0^1 dx\, x^{\alpha-1}(1-x)^{\beta-1}{_2}F_1(a,b;c;zx)
=\frac{\Gamma(\alpha)\Gamma(\beta)}{\Gamma(\alpha+\beta)}
{_3}F_2(a,b,\alpha;c,\alpha+\beta;z).
\label{2F1int}
\end{equation}

\subsection{Integrands with a hypergeometric function}
\label{sec:hyp}

Next, we must consider phase space integrals such as those of
eq.~(\ref{I2def}), where the integrand contains a hypergeometric
function. As is the case for the similar integrals in
refs.~\cite{Anastasiou:2013srw,Anastasiou:2015yha}, we have not found
it possible to obtain a useful closed form analytic result for
arbitrary values of the parameters. However, for a certain subclass of
the parameters, we can indeed find such a result, valid for any
$d$. Let us present this case first.

\subsubsection{The case $\alpha_2=\beta_2=0$}
\label{sec:subclass}

If $\alpha_2$ and $\beta_2$ are both zero, eq.~(\ref{I2def}) reduces
to
\begin{equation}
I_2(\alpha_1,0,\beta_1,0,\gamma_1,\gamma_2,\delta)=
(-2)^{-\alpha_1-\beta_1-\gamma_1-\gamma_2}I(\alpha_1,\beta_1,\gamma_1,
\gamma_2,-\epsilon,4-2\epsilon),
\label{I2reduce}
\end{equation}
where
\begin{align}
I(\alpha_1,\beta_1,\gamma_1,\gamma_2,\delta,a,d)
&=\int d\Phi^{(3)} (p\cdot k_1)^{-\alpha_1}
({\bar{p}\cdot k_1})^{-\beta_1} (p\cdot k_1+ p\cdot k_2)^{-\gamma_1}
(\bar{p}\cdot k_1+\bar{p}_\cdot k_2)^{-\gamma_2}\notag\\
&\quad\times (2k_1\cdot k_2)^\delta
{_2}F_1\left(1,1;a+1;\frac{p\cdot k_1}
{p\cdot k_1+p\cdot k_2}\right).
\label{Idef}
\end{align}
In the centre of mass frame of the two outgoing gluons (see
section~\ref{sec:nohyp}), this becomes
\begin{align}
&I(\alpha_1,\beta_1,\gamma_1,\gamma_2,\delta,a,d)=
2^{2\alpha_1+\gamma_1-\gamma_2}\int d\Phi^{(3)}\,
s_{12}^\delta\,(s-\tilde{t})^{-\alpha_1-\gamma_1}\,A^{-\beta_1-\gamma_2}(1-\cos\theta_1)^{-\alpha_1}
\notag\\
&\quad\times(1-\cos\chi\cos\theta_1
-\sin\chi\sin\theta_1\cos\theta_2)^{-\beta_1}
{_2}F_1\left(1,1;a+1;\frac{1-\cos\theta_1}{2}\right).
\label{Icalc1}
\end{align}
Next, one can use the Mellin-Barnes representation for the
hypergeometric function \newpage
\begin{align}
{_P}F_Q(a_1,\ldots,a_P;b_1,\ldots,b_Q;x)
& =  \int_{-i\infty}^{i\infty}\frac{dw}{2\pi i}
(-x)^w \Gamma(-w) \notag \\ &\times \left[\prod_{i=1}^P\frac{\Gamma(a_i+w)}
{\Gamma(a_i)}\right]\left[\prod_{j=1}^Q\frac{\Gamma(b_i)}
{\Gamma(b_i+w)}\right],
\label{hypMB}
\end{align}
so that eq.~(\ref{Icalc1}) becomes
\begin{align}
&I(\alpha_1,\beta_1,\gamma_1,\gamma_2,\delta,a,d)=
2^{2\alpha_1+\gamma_1-\gamma_2}\int d\Phi^{(3)}\,s_{12}^\delta\,
(s-\tilde{t})^{-\alpha_1-\gamma_1} A^{-\beta_1-\gamma_2}\Gamma(1+a)\notag\\
&\times \int_{-i\infty}^{i\infty}\frac{dw_1}{2\pi i}\,(-1)^{w_1}\,2^{-w_1}
\frac{\Gamma^2(1+w_1)\Gamma(-w_1)}{\Gamma(1+a+w_1)}
(1-\cos\theta_1)^{-(\alpha_1-w_1)}\notag\\
&\times
(1-\cos\chi\cos\theta_1-\sin\chi\sin\theta_1\cos\theta_2)^{-\beta_1}.
\label{MBcalc1}
\end{align}
The angular integrals can be carried out using eq.~(\ref{Ipq}), to get 
\begin{align}
&I(\alpha_1,\beta_1,\gamma_1,\gamma_2,\delta,a,d)=
{\cal N}2^{\alpha_1-\beta_1+\gamma_1-\gamma_2+1}\pi
\frac{\Gamma(d/2-1-\beta_1)\Gamma(1+a)\Gamma(d-3)}{\Gamma^2(d/2-1)}\notag\\
&\times\int_0^1dy\int_0^1dx [y(1-y)]^{d-3}[x(1-x)]^{d/2-2}[1-y(1-z)]^{1-d/2}
s_{12}^\delta\,(s-\tilde{t})^{-\alpha_1-\gamma_1}A^{-\beta_1-\gamma_2}
\notag\\
&\times \int_{-i\infty}^{i\infty}\frac{dw_1}{2\pi i}(-1)^{w_1}
\frac{\Gamma^2(1+w_1)\Gamma(d/2-1-\alpha_1+w_1)\Gamma(-w_1)}
{\Gamma(1+a+w_1)\Gamma(d-2-\alpha_1-\beta_1+w_1)}\notag \\
&\times{_2}F_1\left(\alpha_1-w_1,\beta_1;d/2-1;\cos^2\frac{\chi}{2}\right).
\label{MBcalc2}
\end{align}
At this point, we may expand the integrand in $(1-z)$, taking the
leading power only. After some work, we end up with
\begin{align}
&I(\alpha_1,\beta_1,\gamma_1,\gamma_2,\delta,a,d)
={\cal N}2^{1+\alpha_1+\beta_1+\gamma_1+\gamma_2}\pi
s^{\delta-\alpha_1-\beta_1-\gamma_1-\gamma_2}
(1-z)^{2\delta-\alpha_1-\beta_1-\gamma_1-\gamma_2}\notag\\
&\times\frac{\Gamma(1+a)\Gamma(d/2-1-\beta_1)\Gamma(d-3)}{\Gamma^2(d/2-1)}
\int_0^1 dy\, y^{d-3+\delta-\beta_1-\gamma_2}
(1-y)^{d-3+\delta+\alpha_1-\gamma_1}\notag\\
&\times \int_{-i\infty}^{i\infty}\frac{dw_1}{2\pi i}
(-1)^{w_1}\frac{\Gamma^2(1+w_1)\Gamma(d/2-1-\alpha_1+w_1)\Gamma(-w_1)}
{\Gamma(1+a+w_1)\Gamma(d-2-\alpha_1-\beta_1+w_1)}\notag\\
&\times\int_0^1 dx\, x^{d/2-2+\delta}(1-x)^{d/2-2}
{_2}F_1(\alpha_1-w_1,\beta_1;d/2-1;1-x).
\label{MBcalc3}
\end{align}
The $y$ integral can be carried out immediately in terms of Gamma
functions. The $x$ integral would give a ${_3}F_2$, but then the
remaining Mellin-Barnes integral could be cumbersome. Instead, we can
introduce a second Mellin-Barnes representation, after which the $x$
integral can be carried out in terms of Gamma functions, yielding
\begin{align}
&I(\alpha_1,\beta_1,\gamma_1,\gamma_2,\delta,a,d)
={\cal N}2^{1+\alpha_1+\beta_1+\gamma_1+\gamma_2}\pi
s^{\delta-\alpha_1-\beta_1-\gamma_1-\gamma_2}
(1-z)^{2\delta-\alpha_1-\beta_1-\gamma_1-\gamma_2}\notag\\
&\times\frac{\Gamma(1+a)\Gamma(d/2-1-\beta_1)
\Gamma(d-2+\delta-\beta_1-\gamma_2)\Gamma(d-2+\delta-\alpha_1-\gamma_1)
\Gamma(d/2-1+\delta)}{\Gamma(d/2-1)\Gamma(2d-4+2\delta-\alpha_1-\beta_1
-\gamma_1-\gamma_2)\Gamma(\beta_1)}\notag\\
&\times\Gamma(d-3)\int_{-i\infty}^{i\infty}\frac{dw_1}{2\pi i}
\int_{-i\infty}^{i\infty}\frac{dw_2}{2\pi i}
(-1)^{w_1+w_2}\notag\\
&\times\frac{\Gamma^2(1+w_1)\Gamma(d/2-1-\alpha_1+w_1)
\Gamma(-w_1)\Gamma(\alpha_1-w_1+w_2)\Gamma(\beta_1+w_2)\Gamma(-w_2)}
{\Gamma(\alpha_1-w_1)\Gamma(1+a+w_1)\Gamma(d-2-\alpha_1-\beta_1+w_1)
\Gamma(d-2+\delta+w_2)}.
\label{MBcalc4}
\end{align}
We must now carry out the double Mellin-Barnes integral. However, this
can be done straightforwardly, by recognising the $w_2$ integral as
\begin{align}
&\int_{-i\infty}^{i\infty}\frac{dw_2}{2\pi i}(-1)^{w_2}
\frac{\Gamma(\alpha_1-w_1+w_2)\Gamma(\beta_1+w_2)\Gamma(-w_2)}
{\Gamma(d-2+\delta-w_2)}\notag\\
&\quad=
\frac{\Gamma(\alpha_1-w_1)\Gamma(\beta_1)}
{\Gamma(d-2+\delta)}{_2F_1}(\alpha_1-w_1,\beta_1;
d-2+\delta;1)\notag\\
&\quad=\frac{\Gamma(d-2+\delta)\Gamma(d-2+\delta-\alpha_1-\beta_1+w_1)}
{\Gamma(d-2+\delta-\alpha_1+w_1)\Gamma(d-2+\delta-\beta_1)},
\label{w2int}
\end{align}
where we have used Gauss' identity
\begin{equation}
{_2}F_1(a,b;c;1)=\frac{\Gamma(c)\Gamma(c-a-b)}{\Gamma(c-a)\Gamma(c-b)}.
\label{Gauss}
\end{equation}
At this stage we are left with
\begin{align}
&\hspace{-20pt}I(\alpha_1,\beta_1,\gamma_1,\gamma_2,\delta,a,d)
={\cal N}2^{1+\alpha_1+\beta_1+\gamma_1+\gamma_2}\pi
s^{\delta-\alpha_1-\beta_1-\gamma_1-\gamma_2}
(1-z)^{2\delta-\alpha_1-\beta_1-\gamma_1-\gamma_2}\notag\\
&\times
\frac{\Gamma(d/2-1-\beta-1)\Gamma(d-2+\delta-\beta_1-\gamma_2)
\Gamma(d-2+\delta-\alpha_1-\gamma_1)\Gamma(d/2-1+\delta)}
{\Gamma(d/2-1)\Gamma(2d-4+2\delta-\alpha_1-\beta_1-\gamma_1-\gamma_2)
\Gamma(d-2+\delta-\beta_1)}\notag\\
&\times\Gamma(d-3) \int_{-i\infty}^{i\infty}\frac{dw_1}{2\pi i}
(-1)^{w_1}\notag\\&\times\frac{\Gamma^2(1+w_1)\Gamma(d-2+\delta-\alpha_1-\beta_1+w_1)
\Gamma(d/2-1-\alpha_1+w_1)\Gamma(-w_1)}
{\Gamma(d-2+\delta-\alpha_1+w_1)\Gamma(1+a+w_1)
\Gamma(d-2-\alpha_1-\beta_1+w_1)}.
\label{MBcalc5}
\end{align}
Using eq.~(\ref{hypMB}) we can recognise the $w_1$ integral as
\begin{align}
&\hspace{-10pt}\int_{-i\infty}^{i\infty}\frac{dw_1}{2\pi i}
(-1)^{w_1}\frac{\Gamma^2(1+w_1)\Gamma(d-2+\delta-\alpha_1-\beta_1+w_1)
\Gamma(d/2-1-\alpha_1+w_1)\Gamma(-w_1)}
{\Gamma(d-2+\delta-\alpha_1+w_1)\Gamma(1+a+w_1)
\Gamma(d-2-\alpha_1-\beta_1+w_1)}\notag\\
&\hspace{-10pt}=\frac{\Gamma(d-2+\delta-\alpha_1-\beta_1)\Gamma(d/2-1-\alpha_1)}
{\Gamma(d-2+\delta-\alpha_1)\Gamma(1+a)\Gamma(d-2-\alpha_1-\beta_1)}
\notag\\
&\hspace{-10pt}\times {_4}F_3(1,1,d-2+\delta-\alpha_1-\beta_1,d/2-1-\alpha_1;
d-2+\delta-\alpha_1,1+a,d-2-\alpha_1-\beta_1;1).
\label{w1int}
\end{align}
Putting everything together, we obtain the result of
eq.~(\ref{I2res1}).

\subsection{General parameter values}
\label{sec:gen}

As stated above, for other necessary values of the parameters, we are
not able to find a closed form solution for the integral of
eq.~(\ref{I2def}), valid for any spacetime dimension $d$. Instead, we
may settle for an expansion in the dimensional regularisation
parameter $\epsilon$. To this end, it is useful to use an alternative
phase space parametrisation, as discussed in
refs.~\cite{Anastasiou:2013srw,Anastasiou:2015yha}. We first write
eq.~(\ref{I2def}) as
\begin{equation}
I_2(\alpha_1,\alpha_2,\beta_1,\beta_2,\gamma_1,\gamma_2,\delta)=
(-2)^{-\alpha_1-\alpha_2-\beta_1-\beta_2-\gamma_1-\gamma_2}
J(\alpha_1,\alpha_2,\beta_1,\beta_2,\gamma_1,\gamma_2,\delta,a,d),
\label{I2reduce2}
\end{equation}
where $a=-\epsilon$ and
\begin{align}
&J(\alpha_1,\alpha_2,\beta_1,\beta_2,\gamma_1,\gamma_2,\delta,a,d)
=\int d\Phi^{(3)} (2k_1\cdot k_2)^\delta (p\cdot k_1)^{-\alpha_1}
(p\cdot k_2)^{-\alpha_2}
({\bar{p}\cdot k_1})^{-\beta_1}
({\bar{p}\cdot k_2})^{-\beta_2}\notag\\
&\quad\times (p\cdot k_1+ p\cdot k_2)^{-\gamma_1}
(\bar{p}\cdot k_1+\bar{p}_\cdot k_2)^{-\gamma_2}
{_2}F_1\left(1,1;a+1;\frac{p\cdot k_1}
{p\cdot k_1+p\cdot k_2}\right),
\label{Jdef}
\end{align}
which differs from eq.~(\ref{Idef}) in having arbitrary powers of all
two-particle invariants. Reference~\cite{Anastasiou:2013srw} starts by scaling momenta
according to~\footnote{Our notation $\{p_i\}$ coincides with the
  notation used in ref.~\cite{Anastasiou:2013srw} after the rescaling
  has taken place.}
\begin{equation}
p=\sqrt{s}\,p_1,\quad \bar{p}=\sqrt{s}\,p_2,\quad
k_1=(1-z)\sqrt{s}\,p_3,\quad k_2=(1-z)\sqrt{s}\,p_4.
\label{momscale}
\end{equation}
so that eq.~(\ref{Jdef}) becomes
\begin{align}
&J(\alpha_1,\alpha_2,\beta_1,\beta_2,\gamma_1,\gamma_2,\delta,a,d)=
s^{d-3+\delta-C}
(1-z)^{2d-5+2\delta-C}
\notag\\
&\quad\times\int d\Phi^{(3)} (2p_3\cdot p_4)^\delta (p_1\cdot p_3)^{-\alpha_1}
(p_1\cdot p_4)^{-\alpha_2}
(p_2\cdot p_3)^{-\beta_1}
(p_2\cdot p_4)^{-\beta_2}\notag\\
&\quad\times (p_1\cdot p_3+ p_1\cdot p_4)^{-\gamma_1}
(p_2\cdot p_3+p_2\cdot p_4)^{-\gamma_2}
{_2}F_1\left(1,1;a+1;\frac{p_1\cdot p_3}
{p_1\cdot p_3+p_1\cdot p_4}\right),
\label{Jcalc1}
\end{align}
where 
\begin{displaymath}
C=\alpha_1+\alpha_2+\beta_1+\beta_2+\gamma_1+\gamma_2.
\end{displaymath}
The integral in the second line is now dimensionless. Furthermore, if
one wants the leading behaviour in $(1-z)$, then this has already been
extracted, so that one can set $z=1$ in the integral itself. In
practice this is done by using a particular parametrisation for the
rescaled momenta, and a particular expression for the soft phase
space. The momenta are parametrised in the lab frame, which
immediately implies
\begin{equation}
p_1=\frac{1}{2}\left(1,1,0\ldots\right),\quad
p_2=\frac{1}{2}\left(1,-1,0,\ldots\right).
\label{p12param}
\end{equation}
Furthermore, we can choose to write $p_3$ and $p_4$ in terms of a
d-velocity $\beta_i$:
\begin{equation}
p_i=\frac{E_i}{2}\beta_i,\quad i\in\{3,4\}.
\label{p34param}
\end{equation}
Note that, despite appearances, $E_i$ is dimensionless due to the
rescaling introduced above. Upon substituting eqs.~(\ref{p12param})
and~(\ref{p34param}) into eq.~(\ref{Jcalc1}), the phase space integral
becomes
\begin{align}
2^{C}&\int d\Phi^{(3)}\, s_{34}^\delta\,s_{13}^{-\alpha_1}\,
s_{14}^{-\alpha_2}\,
s_{23}^{-\beta_1}\,s_{24}^{-\beta_2}
\left(s_{13}+s_{14}\right)^{-\gamma_1}
\left(s_{23}+s_{24}\right)^{-\gamma_2}\notag \\ &\times
{_2}F_1\left(1,1;a+1;\frac{s_{13}}{s_{13}+s_{14}}\right),
\label{Jcalc2}
\end{align}
where following ref.~\cite{Anastasiou:2013srw} we have defined 
\begin{equation}
s_{ij}=2p_i\cdot p_j,
\label{sijdef}
\end{equation} where the current notation $s_{12}$ should not be confused with the scale $s_{12}= 2k_1\cdot k_2$  used in the main text. 
At this point one may introduce the Mellin-Barnes representation (see
e.g. ref.~\cite{Smirnov:2004ym})
\begin{align}
{_2}F_1\left(a,b;c;z\right)&=\frac{\Gamma(c)}{\Gamma(a)\Gamma(b)
\Gamma(c-a)\Gamma(c-b)}\notag\\
&\quad\times\int_{-i\infty}^{i\infty}\frac{dz_1}{2\pi i}
\Gamma(a+z_1)\Gamma(b+z_1)\Gamma(c-a-b-z_1)\Gamma(-z_1)(1-z)^{z_1},
\label{hyprep2}
\end{align}
as well as the identity 
\begin{equation}
\frac{1}{(A+B)^\lambda}=\frac{1}{\Gamma(\lambda)}
\int_{-i\infty}^{i\infty}\frac{dz}{2\pi i}\Gamma(-z)\Gamma(\lambda+z)
\frac{A^z}{B^{z+\lambda}},
\label{MBint}
\end{equation}for values of $\lambda>0$,
to rewrite the combinations $(s_{13}+s_{14})$ and
$(s_{23}+s_{24})$. Then, eq.~(\ref{Jcalc2}) assumes the triple
Mellin-Barnes form
\begin{align}
2^{C}&\frac{\Gamma(a+1)}{\Gamma^2(a)\Gamma(\gamma_2)}
\int_{-i\infty}^{i\infty}\frac{dz_1}{2\pi i}
\int_{-i\infty}^{i\infty}\frac{dz_2}{2\pi i}
\int_{-i\infty}^{i\infty}\frac{dz_3}{2\pi i}\notag\\
&\times
\frac{\Gamma^2(1+z_1)\Gamma(a-1-z_1)\Gamma(\gamma_1+z_1+z_2)\Gamma(\gamma_2+z_3)
\Gamma(-z_1)\Gamma(-z_2)\Gamma(-z_3)}{\Gamma(\gamma_1+z_1)}\notag\\
&\times\int d\Phi^{(3)}\,s_{34}^\delta\,
s_{13}^{z_2-\alpha_1}\,s_{14}^{-z_2-\alpha_2-\gamma_1}\,
s_{23}^{z_3-\beta_1}\,s_{24}^{-z_3-\beta_2-\gamma_2}.
\label{Jcalc3}
\end{align}
The phase space integral now has the form of multiple products of
two-particle invariants, thus is of the same form as the integrals
considered in refs.~\cite{Anastasiou:2013srw,Anastasiou:2015yha}. The
invariants can be rewritten using the parametrisation of
eqs.~(\ref{p12param}) and~(\ref{p34param}):
\begin{equation}
s_{1i}=\frac{E_i}{2}\beta_1\cdot \beta_i,\quad
s_{2i}=\frac{E_i}{2}\beta_2\cdot \beta_i,\quad
s_{34}=\frac{E_3\, E_4}{2}\beta_3\cdot \beta_4,\quad i\in\{3,4\}.
\label{sijparam}
\end{equation}
Furthermore, the leading behaviour of the phase space measure as
$z\rightarrow 1$ is given by~\footnote{Given that the soft region
  contributes only at next-to-leading power in $(1-z)$, the leading
  behaviour in $(1-z)$ is sufficient for our purposes.} (see
e.g. ref.~\cite{Anastasiou:2013srw})
\begin{equation}
d\Phi^{(3)}\xrightarrow{z\rightarrow 1}(2\pi)^{3-2d}2^{-2(d-1)}
\delta(1-E_3-E_4)\prod_{i=3}^4 E_i^{d-3} \,dE_i\, d\Omega^{(d-1)}_i,
\label{PS}
\end{equation}
where $d\Omega^{(d-1)}_i$ is the differential solid angle associated with
particle $i$. Using eqs.~(\ref{sijparam}) and~(\ref{PS}) in
eq.~(\ref{Jcalc3}), one may carry out the $E_i$ integrals using
\begin{equation}
\int_0^1 dE_3 \int_0^1 dE_4 \,\delta(1-E_3-E_4)\, 
E_3^{\lambda_3-1}
\,E_4^{\lambda_4-1}=\frac{\Gamma(\lambda_3)\Gamma(\lambda_4)}
{\Gamma(\lambda_3+\lambda_4)},
\label{E34int}
\end{equation}
yielding
\begin{align}
&2^{2C-\delta+5-4d}\pi^{3-2d}\frac{\Gamma(a+1)}{\Gamma^2(a)\Gamma(\gamma_2)
\Gamma(2d-C+2\delta-4)}
\int_{-i\infty}^\infty\frac{dz_1}{2\pi i}
\int_{-i\infty}^\infty\frac{dz_2}{2\pi i}
\int_{-i\infty}^\infty\frac{dz_3}{2\pi i}\notag\\
&\quad \times
\frac{\Gamma^2(1+z_1)\Gamma(a-1-z_1)\Gamma(\gamma_1+z_1+z_2)\Gamma(\gamma_2+z_3)
\Gamma(-z_1)\Gamma(-z_2)\Gamma(-z_3)}{\Gamma(\gamma_1+z_1)}\notag\\
&\quad\times \Gamma(z_2+z_3+d-\alpha_1-\beta_1+\delta-2)
\Gamma(d-z_2-z_3-\alpha_2-\beta_2-\gamma_1-\gamma_2+\delta-2)\notag\\
&\quad\times \int d\Omega^{(d-1)}_3\int d\Omega^{(d-1)}_4\,
(\beta_3\cdot\beta_4)^\delta
(\beta_1\cdot\beta_3)^{z_2-\alpha_1}\,
(\beta_2\cdot\beta_3)^{z_3-\beta_1}\,\notag\\
&\quad\times
(\beta_1\cdot\beta_4)^{-z_2-\alpha_2-\gamma_1}\,
(\beta_2\cdot\beta_4)^{-z_3-\beta_2-\gamma_2}.
\label{Jcalc4}
\end{align}
Next, we must carry out the angular integrals. Given that each
$d$-velocity $\beta_3$ and $\beta_4$ occurs thrice rather than twice,
we can no longer use eq.~(\ref{Ipq}). Unfortunately, there is no known closed form for the
angular integral involving three angular quantities. There is,
however, a triple Mellin-Barnes form~\cite{Somogyi:2011ir} (see also
eq.~(5.17) of ref.~\cite{Anastasiou:2013srw}) in $d=4-2\epsilon$
dimensions:
\begin{align}
&\int d\Omega^{(d-1)}_i (\beta_i\cdot \beta_{j_1})^{-\lambda_1}
(\beta_i\cdot \beta_{j_2})^{-\lambda_2}
(\beta_i\cdot \beta_{j_3})^{-\lambda_3}=
\frac{2^{2-\lambda_1-\lambda_2-\lambda_3-2\epsilon}\pi^{1-\epsilon}}
{\Gamma(\lambda_1)\Gamma(\lambda_2)\Gamma(\lambda_3)
\Gamma(2-\lambda_1-\lambda_2-\lambda_3-2\epsilon)}\notag\\
&\quad\times \int_{-i\infty}^{i\infty}
\frac{dz_4}{2\pi i}\int_{-i\infty}^{i\infty}
\frac{dz_5}{2\pi i}\int_{-i\infty}^{i\infty}
\frac{dz_6}{2\pi i}
\Gamma(-z_4)\Gamma(-z_5)\Gamma(-z_6)\notag\\
&\quad\times\Gamma(\lambda_1+z_4+z_5)\Gamma(\lambda_2+z_4+z_6)
\Gamma(\lambda_3+z_5+z_6)\Gamma(1-\lambda_1-\lambda_2-\lambda_3
-\epsilon-z_4-z_5-z_6)\notag\\
&\quad\times \left(\frac{\beta_{j_1}\cdot\beta_{j_2}}{2}\right)^{z_4}
\left(\frac{\beta_{j_1}\cdot\beta_{j_3}}{2}\right)^{z_5}
\left(\frac{\beta_{j_2}\cdot\beta_{j_3}}{2}\right)^{z_6}.
\label{angint3}
\end{align}
Upon using this result, the remaining integral over the angular
variables of particle 4 can be carried out using eq.~(\ref{Ipq}),
which it is more convenient to write as
\begin{align}
\int d\Omega^{(d-1)}_i (\beta_i\cdot \beta_{j_1})^{-\lambda_1}
(\beta_i\cdot \beta_{j_2})^{-\lambda_2}&=
2^{2-\lambda_1-\lambda_2-2\epsilon}\pi^{1-\epsilon}
\frac{\Gamma(1-\epsilon-\lambda_1)
\Gamma(1-\epsilon-\lambda_2)}{\Gamma(1-\epsilon)\Gamma(2-2\epsilon
-\lambda_1-\lambda_2)}\notag\\
&\quad\times{_2}F_1\left(\lambda_1,\lambda_2;
1-\epsilon;1-\frac{\beta_{j_1}\cdot \beta_{j_2}}{2}\right).
\label{angint2}
\end{align}
Our general phase space integral now has the form of a six-fold
Mellin-Barnes integral, which applies if $\gamma_1$ and $\gamma_2$ are
both non-zero. If either of them is zero, we do not need to apply
eq.~(\ref{MBint}) for the relevant combination of invariants, and thus
we will obtain a lower order Mellin-Barnes integral from the
outset. Our strategy for carrying out an integral for general
$(\alpha_1,\alpha_2,\beta_1,\beta_2,\gamma_1,\gamma_2,\delta,a)$ is as
follows:
\begin{enumerate}
\item For specific parameter values, one should try to reduce the
  five fold MB integral using Barnes' lemmas. We have found that this
  is indeed possible for many integrals.
\item One must shift the contours of the MB integrals, picking up
  residues of poles where appropriate, to extract all singularities in
  $\epsilon$. The output of this procedure is a set of (possibly
  simpler) MB integrals whose integrands can be safely expanded in
  $\epsilon$. To shift the contours, we use the publicly available
  package \texttt{MBResolve}~\cite{Smirnov:2009up}.
\item One can expand the integrands in $\epsilon$, and apply Barnes'
  lemmas where possible to simplify the list of Mellin-Barnes
  integrals. This is done using a combination of the publicly
  available packages \texttt{MB}~\cite{Czakon:2005rk} and
  \texttt{barnesroutines}. At this stage, the output consists of a
  list of (simpler) Mellin-Barnes integrals, some of which will have
  been completely carried out. 
\item Each remaining integral can be carried out in terms of infinite
  sums, for which we use \texttt{MBsums}~\cite{Ochman:2015fho}. The
  resulting sums must then be carried out explicitly, and added
  together. Here, we use the package
  \texttt{xSummer}~\cite{Moch:2005uc}, which itself relies on
  \texttt{FORM}~\cite{Vermaseren:2000nd}~\footnote{We are extremely
    grateful to \"{O}mer G\"{u}rdo\u{g}an for providing an interface
    from \texttt{Mathematica} to \texttt{xSummer}.}. 
\end{enumerate}
All analytic results for the $\epsilon$ expansions of Mellin-Barnes
integrals have been checked numerically using the package \texttt{MB}.
A complication in step 6 is that the individual sums may not converge,
and even the sum of the sums may not converge. In such cases, we
introduce a regulator $x^z$ into the MB integral (where $z$ is the
Mellin variable), before taking the limit $x\rightarrow 1$ having
carried out all sums. An additional possible complication (at step 2)
is that \texttt{MBResolve} may not be able to resolve the
singularities in $\epsilon$. Here one can apply extra regulators to
deal with the problem, as documented in ref.~\cite{Smirnov:2009up}.

Note that the above method will fail if either of the parameters
$(\gamma_1,\gamma_2)$ is negative, given that eq.~(\ref{MBint})
assumes that the left-hand side is a genuine denominator. We indeed
encounter such integrals, with parameter values  (all with
$a=-\epsilon$ and $d=4-2\epsilon$):
\begin{displaymath}
J(1,0,0,1,0,-1,-\epsilon-1,a,d),\,
J(0,0,0,1,1,-1,-\epsilon-1,a,d),\,
J(2,0,0,1,-1,-1,-\epsilon-1,a,d).
\end{displaymath}
Using the simple identities
\begin{equation}
\frac{\bar{p}\cdot(k_1+k_2)}{\bar{p}\cdot k_2}=
\frac{\bar{p}\cdot k_1}{\bar{p}\cdot k_2}+1,\quad
\frac{p\cdot(k_1+k_2)}{p\cdot k_2}=
\frac{p\cdot k_1}{p\cdot k_2}+1,
\label{pfrac}
\end{equation}
we may derive the following relations:
\begin{align}
J(1,0,0,1,0,-1,\delta,a,d)&=
J(1,0,-1,1,0,0,\delta,a,d)+
J(1,0,0,0,0,0,\delta,a,d);\notag\\
J(0,0,1,1,-1,0,\delta,a,d)&=
J(0,0,-1,1,1,0,\delta,a,d)+
J(0,0,0,0,1,0,\delta,a,d);\notag\\
J(2,0,0,1,-1,-1,\delta,a,d)&=
J(1,0,-1,1,0,0,\delta,a,d)+J(1,0,0,0,0,0,\delta,a,d)\notag\\
&\quad+J(2,-1,-1,1,0,0,\delta,a,d)+J(2,-1,0,0,0,0,\delta,a,d).
\label{Jrels}
\end{align}
Integrals on the right-hand side that only involve powers of $p\cdot
k_1$ and / or $\bar{p}\cdot k_1$ can be carried out using the analytic
result of eq.~(\ref{I2res1}). Remaining integrals can be carried out
using the Mellin-Barnes approach outlined in this section. Note,
however, that for the second term in the last line of
eq.~(\ref{Jrels}), it is straightforward to derive a closed form,
valid for any $d$. Starting with
the definition
\begin{equation}
J(2,-1,0,0,0,0,\delta,a,d)=\int d\Phi^{(3)} \frac{(2k_1\cdot k_2)^{\delta}
(p\cdot k_2)}{(p\cdot k_1)^2}{_2}F_1\left(1,1;a+1;\frac{p\cdot k_1}
{p\cdot (k_1+k_2)}\right),
\label{J21calc} 
\end{equation}
we may use the centre of mass frame of the two outgoing gluons
(c.f. section~\ref{sec:nohyp}) to get
\begin{align}
&\hspace{-35pt}J(2,-1,0,0,0,0,\delta,a,d)=\frac{2}{(4\pi)^d}
\frac{s^{d-4+\delta}(1-z)^{2d-6+2\delta}}{\Gamma(d-3)}
\notag\\
&\times\frac{\Gamma(d/2-1+\delta)\Gamma(d/2-1)
\Gamma(d-3+\delta)}{\Gamma(2d-5+2\delta)} \int_0^\pi d\theta_1\int_0^\pi d\theta_2\,
\sin^{d-3}\theta_1\,\sin^{d-4}\theta_2\notag \\ &\times
\left(\frac{1+\cos\theta_1}{2}\right)
\left(\frac{1-\cos\theta_1}{2}\right)^{-2}
{_2}F_1\left(1,1;a+1;\frac{1-\cos\theta_1}{2}\right)
\label{J21calc2}
\end{align}
(n.b. we have already carried out the $x$ and $y$ integrals from
eq.~(\ref{dPS3})). The angular integrals can be carried out by
transforming to
\begin{equation}
u=\frac{1-\cos\theta_1}{2},\quad v=\frac{1-\cos\theta_2}{2},
\label{uvdef}
\end{equation}
from which one finds
\begin{align}
&\int_0^\pi d\theta_1\int_0^\pi d\theta_2\,
\sin^{d-3}\theta_1\,\sin^{d-4}\theta_2\,
\left(\frac{1+\cos\theta_1}{2}\right)
\left(\frac{1-\cos\theta_1}{2}\right)^{-2}
{_2}F_1\left(1,1;a+1;\frac{1-\cos\theta_1}{2}\right)\notag\\
&\hspace{35pt}=2^{2d-7}\int_0^1 dv [v(1-v)]^{(d-5)/2}
\int_0^1 du \,
u^{d/2-4}(1-u)^{d/2-1}
{_2}F_1(1,1;a+1;u)\notag\\
&\hspace{35pt}=2^{2d-7}\frac{\Gamma^2((d-3)/2)\Gamma(d/2-3)\Gamma(d/2)}
{\Gamma^2(d-3)}{_3}F_2(1,1,d/2-3;a+1,d-3;1).
\label{J21calc3}
\end{align}
Putting everything together, one obtains
\begin{align}
&J(2,-1,0,0,0,0,\delta,a,d)=\frac{1}{64\pi^d}
s^{d-4+\delta}(1-z)^{2d-6+2\delta}
\frac{\Gamma(d/2-1+\delta)\Gamma(d/2-1)
\Gamma(d-3+\delta)}
{\Gamma(2d-5+2\delta)\Gamma^3(d-3)}\notag\\
&\hspace{35pt}\times \Gamma^2((d-3)/2)\Gamma(d/2-3)\Gamma(d/2)
\,{_3}F_2(1,1,d/2-3;a+1,d-3;1).
\label{J21res}
\end{align}

\subsection{Results}
\label{app:intresults}

We here collect analytic results, as a Laurent expansion in
$\epsilon$, for the quantities $\hat{I}_2(\alpha_1,\beta_1,\alpha_2,\beta_2,\\\gamma_1,\gamma_2,\delta)$
appearing on the right-hand side of eq.~(\ref{I2hatdef}). Given that
we report only logarithmic terms in $(1-z)$, it is sufficient to
expand up to ${\cal O}(\epsilon)$.
\begin{align}
\hat{I}_2(0,0,1,0,2,0,-\epsilon)&=
\frac{1}{12\epsilon^3}-\frac{5\pi^2}{24\epsilon}-\frac{115\zeta_3}{18}
-\frac{337\pi^4\epsilon}{4320};\notag\\
\hat{I}_2(1,0,1,0,2,1,1-\epsilon)&=
\frac{1}{12\epsilon^3}-\frac{1}{12\epsilon^2}-\frac{1}{\epsilon}\left(
\frac14+\frac{5\pi^2}{24}\right)-\frac34+\frac{11\pi^2}{72}
-\frac{115\zeta_3}{18}\notag\\
&\quad+\epsilon\left(
-\frac94+\frac{11\pi^2}{24}-\frac{337\pi^4}{4320}+\frac{67\zeta_3}{18}
\right);\notag\\
\hat{I}_2(0,0,1,0,1,-1,-1-\epsilon)&=
\frac{5}{24\epsilon^3}-\frac{83\pi^2}{144\epsilon}-\frac{659\zeta_3}
{36}-\frac{173\pi^4\epsilon}{960};\notag\\
\hat{I}_2(1,0,1,0,1,0,-\epsilon)&=
\frac{7}{36\epsilon^3}-\frac{103\pi^2}{216\epsilon}
-\frac{775\zeta_3}{54}-\frac{149\pi^4\epsilon}{864};\notag\\
\hat{I}_2(2,0,1,0,1,1,1-\epsilon)&=
\frac{7}{36\epsilon^3}+\frac{5}{36\epsilon^2}-\frac{1}{\epsilon}
\left(\frac{1}{12}+\frac{103\pi^2}{216}\right)
+\frac{1}{12}-\frac{83\pi^2}{216}-\frac{775\zeta_3}{54}\notag\\
&\quad+\epsilon\left(-\frac{1}{12}-\frac{\pi^2}{72}-\frac{149\pi^4}{864}
-\frac{659\zeta_3}{54}\right);\notag\\
\hat{I}_2(0,0,0,1,2,0,-\epsilon)&=
\frac{1}{16\epsilon^3}-\frac{13\pi^2}{96\epsilon}-\frac{23\zeta_3}{6}
-\frac{107\pi^4\epsilon}{1920};\notag\\
\hat{I}_2(1,0,0,1,1,0,-\epsilon)&=
\frac{11}{48\epsilon^3}-\frac{53\pi^2}{96\epsilon}
-\frac{148\zeta_3}{9}-\frac{727\pi^4\epsilon}{3456};\notag\\
\hat{I}_2(1,0,0,1,2,1,1-\epsilon)&=
\frac{1}{6\epsilon^3}+\frac{1}{12\epsilon^2}+\frac{1}{\epsilon}
\left(\frac14-\frac{5\pi^2}{12}\right)
+\frac34-\frac{19\pi^2}{72}-\frac{227\zeta_3}{18}\notag\\
&\quad
+\epsilon\left(\frac94-\frac{19\pi^2}{24}-\frac{167\pi^4}{1080}
-\frac{157\zeta_3}{18}\right);\notag\\
\hat{I}_2(2,0,0,1,1,1,1-\epsilon)&=
\frac{5}{12\epsilon^2}-\frac{1}{4\epsilon}+\frac14-\frac{77\pi^2}{72}
+\epsilon\left(-\frac14+\frac{17\pi^2}{72}-\frac{295\zeta_3}{9}\right);
\notag\\
\hat{I}_2(1,0,-1,1,0,0,-1-\epsilon)&=
\frac{3}{16\epsilon^3}+\frac{19}{48\epsilon^2}-\frac{1}{\epsilon}
\left(\frac{19}{12}+\frac{149\pi^2}{288}\right)
+\frac{19}{3}-\frac{247\pi^2}{288}-\frac{49\zeta_3}{3}\notag\\
&\quad+\epsilon\left(-\frac{76}{3}+\frac{247\pi^2}{72}-\frac{3137\pi^4}{17280}
-\frac{433\zeta_3}{18}\right);\notag\\
\hat{I}_2(1,0,0,0,0,0,-1-\epsilon)&=
\frac{1}{8\epsilon^3}-\frac{41\pi^2}{144\epsilon}-\frac{33\zeta_3}{4}
-\frac{971\pi^4\epsilon}{8640};\notag\\
\hat{I}_2(0,0,-1,1,1,0,-1-\epsilon)&=
-\frac{1}{24\epsilon^3}+\frac{19}{48\epsilon^2}
+\frac{1}{\epsilon}\left(-\frac{19}{12}+\frac{13\pi^2}{144}\right)
+\frac{19}{3}-\frac{247\pi^2}{288}+\frac{47\zeta_3}{18}\notag\\
&\quad+\epsilon\left(-\frac{76}{3}+\frac{247\pi^2}{72}+\frac{41\pi^4}{960}
-\frac{433\zeta_3}{18}\right);\notag\\
\hat{I}_2(2,0,0,1,0,0,-\epsilon)&=
\frac{11}{48\epsilon^3}+\frac{2}{3\epsilon^2}-\frac{1}{\epsilon}
\left(\frac13+\frac{53\pi^2}{96}\right)
+\frac13-\frac{59\pi^2}{36}-\frac{148\zeta_3}{9}\notag\\
&\quad+\epsilon\left(
-\frac13+\frac{2\pi^2}{3}-\frac{727\pi^4}{3456}-\frac{887\zeta_3}{18}
\right);\notag\\
\hat{I}_2(0,0,0,0,1,0,-1-\epsilon)&=
\frac{1}{24\epsilon^3}-\frac{13\pi^2}{144\epsilon}
-\frac{47\zeta_3}{18}-\frac{41\pi^4\epsilon}{960};\notag\\
\hat{I}_2(2,-1,-1,1,0,0,-1-\epsilon)&=
\frac{5}{16\epsilon^3}+\frac{1}{\epsilon^2}-\frac{1}{\epsilon}\left(
\frac{1}{3}+\frac{77\pi^2}{96}\right)
+\frac13-\frac{95\pi^2}{36}-\frac{295\zeta_3}{12}\notag\\
&\quad+\left(-\frac13+\frac{5\pi^2}{9}-\frac{491\zeta_3}{6}
-\frac{1693\pi^4}{5760}\right);\notag\\
\hat{I}_2(2,-1,0,0,0,0,-1-\epsilon)&=
\frac{1}{4\epsilon^2}-\frac{1}{6\epsilon}+\frac16-\frac{41\pi^2}{72}
+\epsilon\left(-\frac16+\frac{13\pi^2}{36}-\frac{33\zeta_3}{2}
\right).
\label{I2hatresults}
\end{align}

\bibliography{paperv3.bib}


\end{document}